\documentclass[9pt]{article}
\usepackage{amssymb,amsmath,epsfig}
\usepackage{cite}

\begin{document}

\title{\bf Study of Anisotropic Compact Stars with Quintessence Field and Modified Chaplygin Gas in $f(T)$ Gravity}

\author{\bf{Pameli Saha}\thanks{pameli.saha15@gmail.com}~ and
\bf{Ujjal Debnath}\thanks{ujjaldebnath@gmail.com}\\
Department of Mathematics, Indian Institute of Engineering\\
Science and Technology, Shibpur, Howrah-711 103, India.\\}

\date{}
\maketitle

\begin{abstract}
In this work, we get an idea of the existence of compact stars in
the background of $f(T)$ modified gravity where $T$ is a scalar
torsion. We acquire the equations of motion using anisotropic
property within the spherically compact star with electromagnetic
field, quintessence field and modified Chaplygin gas in the
framework of modified $f(T)$ gravity. Then by matching condition,
we derive the unknown constants of our model to obtain many
physical quantities to give a sketch of its nature and also study
anisotropic behavior, energy conditions and stability. Finally, we
estimate the numerical values of mass, surface redshift etc from
our model to compare with the observational data for different
types of compact stars.
\end{abstract}

\sloppy \tableofcontents

\newpage

\section{Introduction:}

Recently, compact stars, the most elemental objects of the galaxies,  acquire much attention to the researchers to study their ages, structures and evolutions in cosmology as well as astrophysics. After the stellar death, the residue portion is formed as compact stars which can be classified into white dwarfs, neutron stars, strange stars and black holes. Compact star is very densed object i.e.,  posses massive mass and smaller radius as compared to the ordinary star. In astrophysics, the study of neutron star and the strange star motivates the researchers to explore their features and structures very much. Massive neutron star again collapses into black hole but lower mass neutron star converts into quark star. Basically, neutron star consists of neutrons whereas strange star is made up with quarks or strange matters. After the discovery of neutrons \cite{BWZF34}, the researchers have first imagined about the presence of neutron star. Then Hewish et al. \cite{HABSJPJDH68} has confirmed this prediction by observation of pulsars (considered as rotating neutron stars later) like Her X-$1$, $4$U $1820-30$, RXJ $1856-37$ and SAX J $1808.4-3658$. In 1916, Karl Schwarzschild \cite{1SK16,2SK16} has first given the interior stellar solution which must be matched with the exterior solution. We have noticed for a long time that the isotropic fluid (having equal radial pressure ($p_{r}$) and transversal pressure ($p_{t}$)) is considered in the core of the stellar objects to study stellar structure and stellar evolution. In 1972, Ruderman \cite{RR72} for the first time observed that the interior geometry of the nuclear matter with a density of order $10^{15} gm/cc$ posses anisotropic behavior (having $p_{r}\neq p_{t}$). This nature may come from different sources: the existence of solid core, in presence of type P superfluid, phase transition, rotation, magnetic field, mixture of two fluids, viscosity etc. Herrera et al. \cite{HLSNO97} have given a review to analyze local anisotropic nature for self gravitating systems. A stable structure of stellar objects has been found in the context of anisotropic nature by Hossein et al. \cite{HSKM12}. Kalam et al. \cite{KMRFHSKM14} have investigated anisotropic neutron star with quintessence dark energy. A new exact solution for compact star has been discovered by Paul et al. \cite{PBCDR14} to preserve the hydrostatic equilibrium. Many astrophysical phenomena of quark star and neutron star have been discussed in ref. \cite{WE84,CKSDZGLT98}. Rahaman et al. \cite{RFCKKPKF14} have observed the existence of strange star using MIT bag model to study mass and redshift functions. Again, a stable anisotropic quintessence strange star model has been proposed by Bhar \cite{BP14}. Murad \cite{MMH16} has also studied anisotropic charged strange star with MIT bag model to find out the radial pressure and energy density. A stability of strange star with the influence of anisotropic context using MIT bag model has been investigated by Arba$\tilde{n}$il and Malheiro \cite{AJDV16}.\\

After publishing of Einstein's General Relativity (GR) in 1915, our mysterious universe has been sketched more clearly to people. Recently, we have come to know from many observational evidences like Large Scale Structure (LSS), Cosmic Microwave Background (CMB) radiation etc. that our universe is expanding with acceleration which asserts that the cosmic expansion is going due to some peculiar source of energy having a massive negative pressure, known as Dark Energy (DE). Similarly, there exists also a mysterious component, known as Dark Matter (DM). A telescope is unable to detect dark matter but its gravitational effects on visible matter and gravitational lensing of the background radiation are giving us the evidence of its existence. On the other hand the Equation of State (EoS) of DE is given by $w=p/\rho$ where $w$ is called the EoS parameter lying in the range $w<-1/3$. If $-1<w<-1/3$ then it is referred as Quintessence DE and if $w<-1$ then another peculiar DE viz., as Phantom DE producing Big Rip Singularity and violating Null Energy Condition (NEC) also. The phantom DE has been discussed in many references \cite{RFKMSMGK06,LFSN05,1LFSN05}. In particular, $w=-1$ gives the EoS of ``Gravastar", gravitationally vacuum condense star \cite{1MPME01,1MPME04,1UAARF11,1RFRSUAA12,1RFUAARSIS12,1BP14}.\\

However, GR is not sufficiently enough to describe our present universe from theoretical as well as physical point of view. So, we need an alternative theories to GR to present the scenario of DE and DM. We know from the references that if the torsion scalar $T$ is instead of Ricci scalar $R$ in Einstein-Hilbert action then the obtained equations of motion of this theory of gravity, known as Teleparallel Equivalent of General Relativity (TEGR) \cite{ARPJG13}, are equivalent to those of GR. If generalization of GR is possible by inserting $f(R)$ or $f(T)$ instead of $R$ or $T$ respectively then our accelerating expansion of our universe can be explained where $f(R)$ or $f(T)$ are arbitrary nonlinear functions of $R$ or $T$. It is seen that fourth order differential equations arrive in the case of $f(R)$ gravity whereas $f(T)$ gravity gives second order differential equations in the tetrad field like GR. So, the later approach is more convenient than the former one. In the background space-time, there has nonzero torsion with no curvature. According to Einstein, this is the definition of space-time to relate gravitation with tetrad and torsion. Many authors have revealed wide interest in \cite{WCM06,BGRFR09,LEV10,MR11,CYFCS16,YRJ11,BKGCQ11,WPYH11,CSCVF14,NSOSDOVK17,FGSJLRML16,ILRNRML15,QJZCS17}.\\

In theoretical astrophysics, Dent \cite{Dent} has derived the solution of BTZ black hole in $f(T)$ version in 3-dimensions. Later, first violation of black hole thermodynamics in $f(T)$ gravity has been seen through violation of Lorentz invariance \cite{Rong-Xin}. For the existence of astrophysical stars in $f(T)$ theory, the physical conditions have been investigated \cite{Daouda} with very much attention after acquiring a large group of static perfect fluid solutions \cite{Boehmer}. Some static solutions for spherically symmetric case with charged source in $f(T)$ theory has been obtained \cite{Wang}. Capozziello et. al \cite{1CSGPA13} have observed the removal of singularity of the exact black hole by $f(T)$ gravity instead of $f(R)$ gravity in D dimensions. Sharif and Rani \cite{1SMRS13} have studied wormhole under $f(T)$ gravity. They have studied static wormhole solution in $f(T)$ gravity to investigate energy conditions \cite{1SMRS14} and also proved that this solution exists by violating energy conditions for charged noncommutative wormhole \cite{2SMRS14,3SMRS14}.\\

The study of anisotropic stars in the background of General
Relativity (GR) and modified gravity has drawn much interest to
many researchers \cite{Abbas1,Abbas2,Abbas3,Abbas4,Abbas5,Mak}.
Krori-Barua (KB) metric \cite{Krori,Kalam1} is the most useful
metric to discuss the compact star models. Abbas and his
collaborations \cite{Abbas6,Abbas7,AG15,Abbas9,Abbas10} have also
studied this type of stars in GR, $f(R)$, $f(G)$ and $f(T)$ with
the help of (KB) metric. Abbas et al. \cite{Abbas9} have
investigated anisotropic strange star taking $p=\alpha\rho$ where
$0<\alpha<1$ which plays as a quintessence dark energy model. A
study of strange star with MIT bag model in the framework of
$f(T)$ gravity has also been analyzed by Abbas et al.
\cite{AGQSJA15}. Recently, many authors have studied anisotropic
stars in GR and different types of modified gravity with (KB)
metric
\cite{SMWA18,1MSKBA19,2MSKOFT19,3MSKOFT19,5AGSMR19,PAKKJ19}. Saha
and Debnath \cite{SPDU18} have considered a metric where the
unknown functions $a(r)$ and $b(r)$ have been taken some different
from KB metric to investigate anisotropic stars in $f(T)$ gravity
with modified Chaplygin gas. This model also reveals as a
quintessence dark energy model like the previous cases. In this
work, our motivation is to explore anisotropic compact stars in
the frame-work of $f(T)$ gravity with diagonal tetrad in presence
of electric field and quintessence field along transversal
direction with modified Chaplygin gas by considering a metric like
the previous work \cite{SPDU18} where we take the general forms of
$a(r)$ and $b(r)$ instead of particular forms. In section 2, an
introduction of $f(T)$ gravity is given and we take anisotropic
fluid with quintessence field along transversal direction. In
section 3, we investigate anisotropic compact star with
electromagnetic field and quintessence field in $f(T)$ gravity by
taking modified Chaplygin gas. We calculate all physical
quantities of our proposed model. In section 4, we apply matching
of two metrics to find out the unknown constants of our model. We
observe the nature of our model by plotting some figures. Section
5 gives anisotropic behaviour. We make stability analysis and
verify the energy conditions of our model. Again, we calculate the
mass function, compactness and surface redshift function from our
model to make comparison with observational data. In section 6, we
deliver the conclusions of the work.

\section{An Introduction of $f(T)$ Gravity, Electromagnetic Field and Quintessence Field}

First we review the formulation of $f(T)$ gravity in the concept
of tetrad formalism. We assume the general form of space-time
metric is in the form
\begin{equation}
ds^{2}=g_{\mu\nu}dx^{\mu}dx^{\nu}.
\end{equation}
In the tetrad formalism, the above metric can be written as
\begin{equation}
ds^{2}=\eta_{ij}\theta^{i}\theta^{j}
\end{equation}
where
\begin{equation}
dx^{\mu}=e^{\mu}_{i}\theta^{i}~,~\eta_{ij}=diag[-1,1,1,1]~,~e^{\mu}_{i}e^{\nu}_{j}=\delta^{\mu}_{\nu}.
\end{equation}
The non-vanishing Christoffel symbols are
\begin{equation}
\Gamma^{\alpha}_{\mu\nu}=e^{\alpha}_{i}\partial_{\nu}e^{i}_{\mu}=-e^{i}_{\mu}\partial_{\nu}e^{\alpha}_{i}.
\end{equation}
The torsion and the con-torsion tensor can be defined as follows
\cite{AGQSJA15}:
\begin{equation}\label{1}
T_{\mu\nu}^{\alpha}=\Gamma_{\nu\mu}^{\alpha}-\Gamma_{\mu\nu}^{\alpha}=e_{i}^{\alpha}(\partial_{\mu}e_{\nu}^{i}-\partial_{\nu}e_{\mu}^{i}),
\end{equation}

\begin{equation}\label{2}
K_{\alpha}^{\mu\nu}=-\frac{1}{2}(T_{\alpha}^{\mu\nu}-T_{\alpha}^{\nu\mu}-T_{\alpha}^{\mu\nu}).
\end{equation}

The tensor $S_{\alpha}^{\mu\nu}$ are defined as in the form
\begin{equation}\label{3}
S_{\alpha}^{\mu\nu}=\frac{1}{2}(K_{\alpha}^{\mu\nu}+\delta_{\alpha}^{\mu}T_{\beta}^{\beta\nu}-\delta_{\alpha}^{\nu}T_{\beta}^{\beta\mu}).
\end{equation}

The torsion scalar is defined as follows
\begin{equation}\label{4}
T=T_{\mu\nu}^{\alpha}S_{\alpha}^{\mu\nu}.
\end{equation}

Now, the modified teleparallel action is defined as follows \cite{STPFV10,1BKCSNODS12}
\begin{equation}\label{5}
S=\int d^{4}xe\Big[\frac{1}{16\pi}f(T)+\textit{L}_{Matter}(\Phi_{A})\Big]
\end{equation}
where $G=c=1$ and $\textit{L}_{Matter}(\Phi_{A})$ is the matter Lagrangian.\\

Now, we consider our model containing quintessence field and electromagnetic field along with anisotropic pressure. So, the Einstein equations can taken as
\begin{equation}\label{6}
G_{\mu\nu}=8\pi
G(T_{\mu\nu}^{Matter}+T_{\mu\nu}^{q}+T_{\mu\nu}^{EM}).
\end{equation}
Here, the ordinary matter corresponding to anisotropic fluid has energy-momentum tensor as
\begin{equation}\label{7}
T_{\mu\nu}^{Matter}=(\rho+p_{t})u_{\mu}u_{\nu}-p_{t}g_{\mu\nu}+(p_{r}-p_{t})v_{\mu}v_{\nu}
\end{equation}
where $u^{\mu}$ is the four-velocity and $v^{\mu}$ radial four
vector satisfying $u_{\mu}u^{\mu}=1,~v_{\mu}v^{\mu}=-1$ and
$u_{\mu}v^{\mu}=0$. Here $\rho$ the energy density, $p_{r}$ is the
radial pressure and $p_{t}$ is transverse pressure. Also
$T_{\mu\nu}^{q}$ is the energy momentum tensor of quintessence
field having energy density $\rho_{q}$ and equation of state
parameter $w_{q}$ ($-1<w_{q}<-1/3$). According to Kiselev
\cite{KVV03}, the components of this tensor require to satisfy
additivity and linearity. Taking different signatures as in line
elements, the components are given by
\begin{equation}\label{8}
T_{t}^{t}=T_{r}^{r}=-\rho_{q},
\end{equation}
\begin{equation}\label{9}
T_{\theta}^{\theta}=T_{\phi}^{\phi}=\frac{1}{2}(3w_{q}+1)\rho_{q}.
\end{equation}\\
Further, the energy momentum tensor for electromagnetic field is given by
\begin{equation}\label{10}
E_{\mu\nu}^{EM}=\frac{1}{4\pi}(g^{\delta\omega}F_{\mu\delta}F_{\omega\nu}-\frac{1}{4}g_{\mu\nu}F_{\delta\omega}F^{\delta\omega})
\end{equation}
where $F_{\mu\nu}$ is the Maxwell field tensor defined as
$F_{\mu\nu}=\Phi_{\nu,\mu}-\Phi_{\mu,\nu}$ and $\Phi_{\mu}$ is the
four potential. The corresponding Maxwell electromagnetic field
equations are
\begin{equation}
(\sqrt{-g}~F^{\mu\nu}),_{\nu}=4\pi
J^{\mu}\sqrt{-g}~,~F_{[\mu\nu,\delta]}=0
\end{equation}
where $J^{\mu}$ is the current four-vector satisfying
$J^{\mu}=\sigma u^{\mu}$, the parameter $\sigma$ is the charge
density.

\section{Anisotropic Compact Star with Electromagnetic Field and Quintessence Field in $f(T)$ Gravity:}

Let us assume the spherically symmetrical metric for the interior space-time solution as \cite{KKDBJ75}
\begin{equation}\label{12}
ds^{2}=-e^{a(r)}dt^{2}+e^{b(r)}dr^{2}+r^{2}(d\theta^{2}+\sin^{2}\theta
d\phi^{2}).
\end{equation}
Here, we take $a(r)$ and $b(r)$ in the following forms:
\begin{equation}\label{13}
a(r)=Br^{\alpha}+Cr^{\beta}~~,~~b(r)=Ar^{\gamma}
\end{equation}
where $A$, $B$ and $C$ are arbitrary constants. Here, $\alpha\geq2$, $\beta\geq2$ and $\gamma\geq2$ are constants but $\alpha\neq\beta$. For $\alpha=2$, $\beta=\gamma=3$, the unknown functions $a(r)$ and $b(r)$ reduce to the forms in which anisotropic quintessence star has been studied in \cite{SPDU18}. For this metric, the torsion scalar $T$ and its derivative are given as \cite{AGQSJA15}
\begin{equation}\label{14}
T(r)=\frac{2e^{-b}}{r}\Big(a'+\frac{1}{r}\Big),
\end{equation}
\begin{equation}\label{15}
T'(r)=\frac{2e^{-b}}{r}\Big\{a''-\frac{a'}{r}-b'\Big(a'+\frac{1}{r}\Big)-\frac{2}{r^{2}}\Big\}
\end{equation}
where the prime $'$ denotes the derivative with respect to the radial coordinate $r$.\\

Now, the equations of motion for anisotropic fluid along with
quintessence field and magnetic field in the framework of $f(T)$
gravity are as follows \cite{AGQSJA15}:
\begin{equation}\label{16}
-\frac{e^{-b}}{r}T'f_{TT}+\frac{f}{2}-\Big\{T-\frac{1}{r^{2}}-\frac{e^{-b}}{r}(a'+b')\Big\}f_{T}=8\pi(\rho+\rho_{q})+E^{2},
\end{equation}
\begin{equation}\label{17}
(T-\frac{1}{r^{2}})f_{T}-\frac{f}{2}=8\pi(p_{r}-\rho_{q})-E^{2},
\end{equation}
\begin{equation}\label{18}
e^{-b}\Big(\frac{a'}{2}+\frac{1}{r}\Big)T'f_{TT}+\Big[\frac{T}{2}+e^{-b}\Big\{\frac{a''}{2}+\Big(\frac{a'}{4}+\frac{1}{2r}\Big)(a'-b')\Big\}\Big]f_{T}-\frac{f}{2}=8\pi\Big\{p_{t}+\frac{1}{2}(3w_{q}+1)\rho_{q}\Big\}+E^{2},
\end{equation}
\begin{equation}\label{19}
\frac{e^{-\frac{b}{2}}\cot\theta}{2r^{2}}T'f_{TT}=0,
\end{equation}
\begin{equation}\label{20}
E(r)=\frac{1}{r^{2}}\int_{0}^{r}4\pi r^{2}\sigma(r)e^{\frac{b}{2}}dr=\frac{q(r)}{r^{2}}
\end{equation}
where $q(r)$ is the total charge within a sphere of radius $r$.\\

Here, we take the total charge $q(r)$ as in the power law form:
\begin{equation}\label{21}
q(r)=q_{0}r^{m}
\end{equation}
where $q_{0}>0$, $m>0$. Here we consider the fluid source behaves
as modified Chaplygin gas (MCG) whose equation of state is
\cite{BHB12}
\begin{equation}\label{22}
p_{r}=\xi\rho-\frac{\zeta}{\rho^{\alpha'}}
\end{equation}
where $\xi$, $\alpha'$ and $\zeta$ are free parameters of our model.\\

Now, solving equation (23) we get (assuming $T'\ne 0$)
\begin{equation}\label{23}
f(T)=\beta_{1}T+\beta_{2},
\end{equation}
where $\beta_{1}$ and $\beta_{2}$ being integration constants and
we assume $\beta_{2}=0$ for simple case.\\

Now from equations (14), (15), (17), (18), (22) and (23) (taking
$\alpha'=1$) we obtain the equation in $\rho$:
\begin{equation}\label{24}
8\pi r^{2}\rho^{2}(1+\xi)-\beta_{1}e^{-Ar^{\gamma}}\rho(\alpha Br^{\alpha}+\beta Cr^{\beta}+\gamma Ar^{\gamma})-8\pi r^{2}\zeta=0.
\end{equation}
Solving this equation, we get the value of energy density as
\begin{equation}\label{25}
\rho=\frac{\beta_{1}e^{-Ar^{\gamma}}(\alpha Br^{\alpha}+\beta Cr^{\beta}+\gamma Ar^{\gamma})+\sqrt{\beta_{1}^{2}e^{-2Ar^{\gamma}}(\alpha Br^{\alpha}+\beta Cr^{\beta}+\gamma Ar^{\gamma})^{2}+256\pi^{2}r^{4}\zeta(1+\xi)}}{16\pi r^{2}(1+\xi)}.
\end{equation}

Then we found the expressions of the radial pressure, transverse
pressure and density for quintessence field as

\begin{eqnarray*}
p_{r}=\frac{\xi\beta_{1}e^{-Ar^{\gamma}}(\alpha Br^{\alpha}+\beta Cr^{\beta}+\gamma Ar^{\gamma})+\xi\sqrt{\beta_{1}^{2}e^{-2Ar^{\gamma}}(\alpha Br^{\alpha}+\beta Cr^{\beta}+\gamma Ar^{\gamma})^{2}+256\pi^{2}r^{4}\zeta(1+\xi)}}{16\pi r^{2}(1+\xi)}
\end{eqnarray*}
\begin{equation}\label{26}
-\frac{16\pi\zeta r^{2}(1+\xi)}{\beta_{1}e^{-Ar^{\gamma}}(\alpha Br^{\alpha}+\beta Cr^{\beta}+\gamma Ar^{\gamma})+\sqrt{\beta_{1}^{2}e^{-2Ar^{\gamma}}(\alpha Br^{\alpha}+\beta Cr^{\beta}+\gamma Ar^{\gamma})^{2}+256\pi^{2}r^{4}\zeta(1+\xi)}}.
\end{equation}

\begin{eqnarray*}
p_{t}=\frac{1}{8\pi}\Big[\beta_{1}e^{-Ar^{\gamma}}\Big\{\frac{1}{2}(\alpha(\alpha-1)Br^{\alpha-2}+\beta(\beta-1)Cr^{\beta-2})+\frac{1}{4}(\alpha
Br^{\alpha-1}+\beta Cr^{\beta-1})(\alpha Br^{\alpha-1}+\beta
Cr^{\beta-1}
\end{eqnarray*}
\begin{eqnarray*}
-\gamma Ar^{\gamma-1})+\frac{1}{2}(\alpha Br^{\alpha-2}+\beta
Cr^{\beta-2}-\gamma
Ar^{\gamma-2})\Big\}-\frac{\beta_{2}}{2}-\frac{q^{2}}{r^{4}}\Big]-\frac{1}{2}(3w_{q}+1)\frac{1}{8\pi}\Big\{\frac{\beta_{1}e^{-Ar^{\gamma}}}{r^{2}}(\gamma
Ar^{\gamma}-1)+\frac{\beta_{1}}{r^{2}}+\frac{\beta_{2}}{2}
\end{eqnarray*}
\begin{equation}\label{27}
-\frac{q^{2}}{r^{4}}\Big\}-\frac{\beta_{1}e^{-Ar^{\gamma}}(\alpha
Br^{\alpha}+\beta Cr^{\beta}+\gamma
Ar^{\gamma})+\sqrt{\beta_{1}^{2}e^{-2Ar^{\gamma}}(\alpha
Br^{\alpha}+\beta Cr^{\beta}+\gamma
Ar^{\gamma})^{2}+256\pi^{2}r^{4}\zeta(1+\xi)}}{16\pi
r^{2}(1+\xi)}.
\end{equation}

\begin{eqnarray*}
\rho_{q}=\frac{1}{8\pi}\Big\{\frac{\beta_{1}e^{-Ar^{\gamma}}}{r^{2}}(\gamma Ar^{\gamma}-1)+\frac{\beta_{1}}{r^{2}}+\frac{\beta_{2}}{2}-\frac{q^{2}}{r^{4}}\Big\}-\frac{\beta_{1}e^{-Ar^{\gamma}}(\alpha Br^{\alpha}+\beta Cr^{\beta}+\gamma Ar^{\gamma})}{16\pi r^{2}(1+\xi)}
\end{eqnarray*}
\begin{equation}\label{28}
+\frac{\sqrt{\beta_{1}^{2}e^{-2Ar^{\gamma}}(\alpha Br^{\alpha}+\beta Cr^{\beta}+\gamma Ar^{\gamma})^{2}+256\pi^{2}r^{4}\zeta(1+\xi)}}{16\pi r^{2}(1+\xi)}.
\end{equation}

\textbf{The equation of state parameters $w_{r}$ and $w_{t}$ along radial and transversal directions of our model are given by
\begin{equation}
w_{r}=\frac{p_{r}}{\rho}=\xi-\Big\{\frac{16\pi\zeta r^{2}(1+\xi)}{\beta_{1}e^{-Ar^{\gamma}}(\alpha Br^{\alpha}+\beta Cr^{\beta}+\gamma Ar^{\gamma})+\sqrt{\beta_{1}^{2}e^{-2Ar^{\gamma}}(\alpha Br^{\alpha}+\beta Cr^{\beta}+\gamma Ar^{\gamma})^{2}+256\pi^{2}r^{4}\zeta(1+\xi)}}\Big\}^{2}
\end{equation}
and
\begin{equation}
w_{t}=\frac{p_{t}}{\rho}
\end{equation}
where $p_{t}$ and $\rho$ are given by equations (31) and (29).}\\

The above solutions are viable if $\xi\ne -1$.

\section{Matching Conditions:}

By matching condition, many researchers have compared the exterior
solution with the interior solution
\cite{1BP14,Abbas7,Abbas9,AGQSJA15}. We make correspondence
between our interior solution and the exterior solution evoked by
Reissner-Nordstrm metric whose line element is
\begin{equation}\label{29}
ds^{2}=-\Big(1-\frac{2M}{r}+\frac{Q^{2}}{r^{2}}\Big)dt^{2}+\Big(1-\frac{2M}{r}+\frac{Q^{2}}{r^{2}}\Big)^{-1}dr^{2}
+r^{2}(d\theta^{2}+\sin^{2}\theta d\phi^{2}).
\end{equation}
where $m$, $r$ and $Q$ are mass, radius and charge respectively.
We assume the boundary of interior and exterior regions occurs at
$r=R$. So on the boundary, we have $m(r)=M$ and
\begin{equation}\label{30}
g_{tt}^{-}=g_{tt}^{+},~~~~~g_{rr}^{-}=g_{rr}^{+},~~~~~\frac{\partial g_{tt}^{-}}{\partial r}=\frac{\partial g_{tt}^{+}}{\partial r},
\end{equation}
where $-$ and $+$ indicate to interior and exterior solutions.
Now, using \textbf{(36)} and the metrics (16), \textbf{(35)}, we can obtain
\begin{equation}\label{31}
\left.
\begin{array}{ll}
A=-\frac{1}{R^{\gamma}}\ln\Big(1-\frac{2M}{R}+\frac{Q^{2}}{R^{2}}\Big),\\
B=\frac{1}{R^{\alpha}(\alpha-\beta)}\Big[-\beta\ln\Big(1-\frac{2M}{R}+\frac{Q^{2}}{R^{2}}\Big)+2\Big(\frac{M}{R}-\frac{Q^{2}}{R^{2}}\Big)\Big(1-\frac{2M}{R}+\frac{Q^{2}}{R^{2}}\Big)^{-1}\Big],\\
C=\frac{1}{R^{\beta}(\alpha-\beta)}\Big[\alpha\ln\Big(1-\frac{2M}{R}+\frac{Q^{2}}{R^{2}}\Big)-2\Big(\frac{M}{R}-\frac{Q^{2}}{R^{2}}\Big)\Big(1-\frac{2M}{R}+\frac{Q^{2}}{R^{2}}\Big)^{-1}\Big]
\end{array}
\right\}
\end{equation}

We consider three different compact stars as $Vela~X$-$1~(CS1)$,
$SAXJ1808.4$-$3658~(CS2)$ and $4U1820$-$30~(CS3)$ with their
masses and radii in Table 1. With respect to Table 1, we take
three sets of values of $\alpha$, $\beta$ and $\gamma$ to get
three different sets of numerical values of $A$, $B$
and $C$ for $CS1$, $CS2$ and $CS3$ in Tables 2-4 respectively.\\

\begin{table}[h]
\centering
\begin{tabular}{|l|l|l|l|l|}
\hline
$Compact~Stars$ & $M(M_{\odot})$ & $R(Km)$ & $\mu_{M}=\frac{M}{R}$ & $\mu_{c}=\frac{Q^{2}}{R^{2}}$ \\
\hline
$Vela~X$-$1~(CS1)$ & 1.77 & 9.56 & 0.273091 & 0.0133624 \\
\hline
$SAXJ1808.4$-$3658~(CS2)$ & 1.435 & 7.07 & 0.299 & 0.0266898 \\
\hline
$4U1820$-$30~(CS3)$ & 2.25 & 10 & 0.332 & 0.0133208 \\
\hline
\end{tabular}
\caption{Different compact stars are taken with their masses and radii.}
\end{table}

\begin{table}[h]
\centering
\begin{tabular}{|l|l|l|l|}
\hline
$Compact~Stars$ & $A (Km^{-2})$ & $B (Km^{-2})$ & $C (Km^{-2})$ \\
\hline
$Vela~X$-$1~(CS1)$ & 0.0008710312 & -0.037147216 & 0.003014661038 \\
\hline
$SAXJ1808.4$-$3658~(CS2)$ & 0.0023968248 & -0.07625293 & 0.008388596829 \\
\hline
$4U1820$-$30~(CS3)$ & 0.0010517646 & -0.049798583 & 0.003928093861 \\
\hline
\end{tabular}
\caption{The values of $A$, $B$ and $C$ have been obtained with
$\alpha=2$, $\beta=\gamma=3$ from Table 1 using equation \textbf{(37)}.}
\end{table}

\begin{table}[h]
\centering
\begin{tabular}{|l|l|l|l|}
\hline
$Compact~Stars$ & $A (Km^{-2})$ & $B (Km^{-2})$ & $C (Km^{-2})$ \\
\hline
$Vela~X$-$1~(CS1)$ & 0.00009111205 & 0.0074533913 & -0.037147217 \\
\hline
$SAXJ1808.4$-$3658~(CS2)$ & 0.00033901341 & 0.0083885968 & -0.076252931 \\
\hline
$4U1820$-$30~(CS3)$ & 0.00010517646 & 0.0039280939 & -0.049798584 \\
\hline
\end{tabular}
\caption{The values of $A$, $B$ and $C$ have been obtained with
$\alpha=3$, $\beta=2$ and $\gamma=4$ from Table 1 using equation
\textbf{(37)}.}
\end{table}

\begin{table}[h]
\centering
\begin{tabular}{|l|l|l|l|}
\hline
$Compact~Stars$ & $A (Km^{-2})$ & $B (Km^{-2})$ & $C (Km^{-2})$ \\
\hline
$Vela~X$-$1~(CS1)$ & 0.00087103116 & 0.00015767056 & -0.022737137 \\
\hline
$SAXJ1808.4$-$3658~(CS2)$ & 0.0023968248 & 0.00059325296 & -0.046599241 \\
\hline
$4U1820$-$30~(CS3)$ & 0.0010517646 & 0.000196404693 & -0.030158115 \\
\hline
\end{tabular}
\caption{The values of $A$, $B$ and $C$ have been obtained with
$\alpha=4$, $\beta=2$ and $\gamma=3$ from Table 1 using equation
\textbf{(37)}.}
\end{table}

\begin{figure}

\includegraphics[height=1.35in]{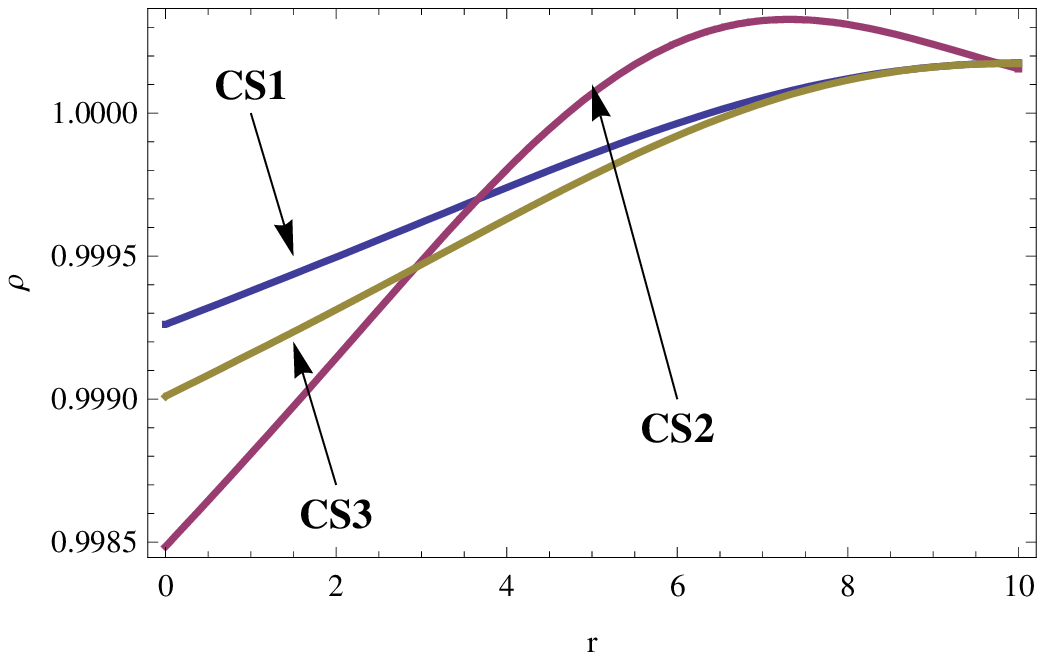}~~
\includegraphics[height=1.35in]{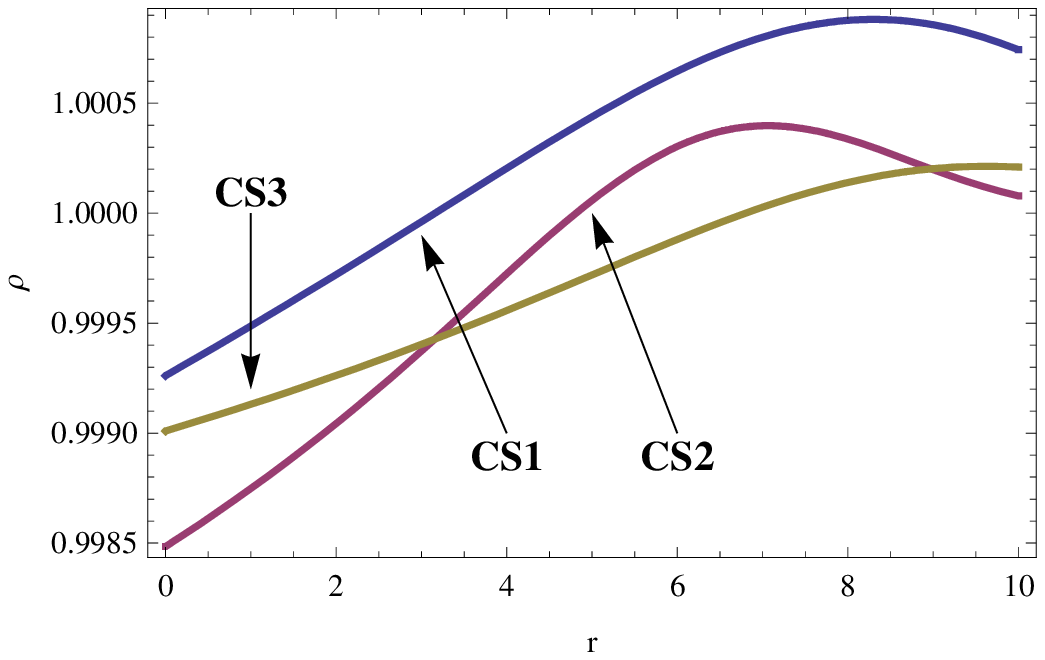}~~
\includegraphics[height=1.35in]{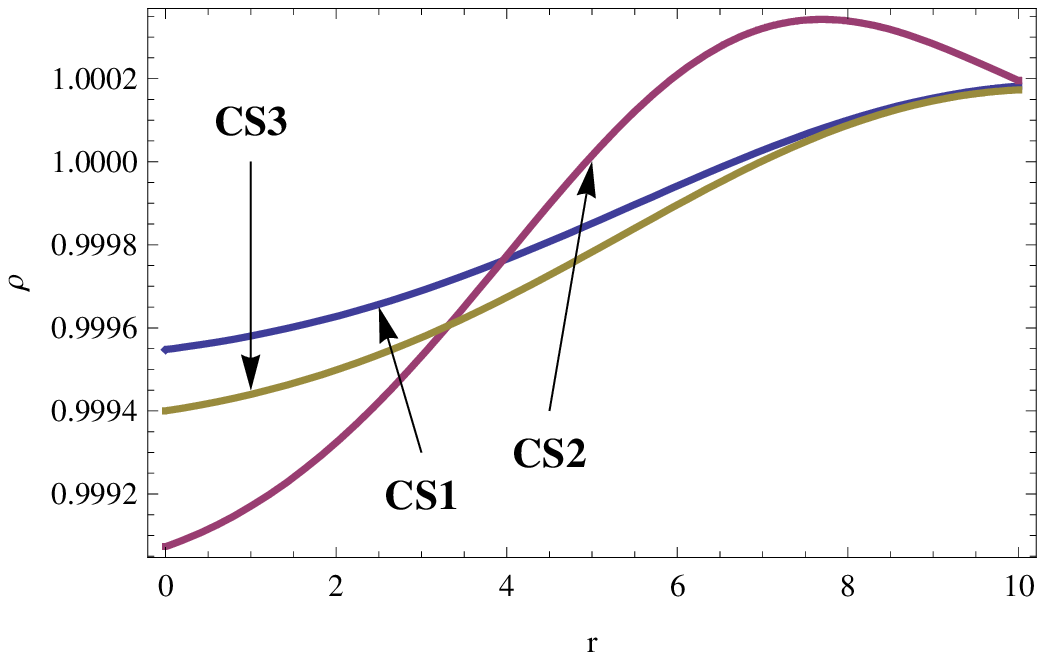}\\
\vspace{2 mm}
\textbf{Fig.1} Variations of energy density \textbf{$\rho$ ($MeV/fm^{3}$)} versus $r$ ($km$) with the numerical values of $A$, $B$ and $C$ from Table 2, Table 3 and Table 4 respectively.\\
\vspace{6 mm}

\includegraphics[height=1.35in]{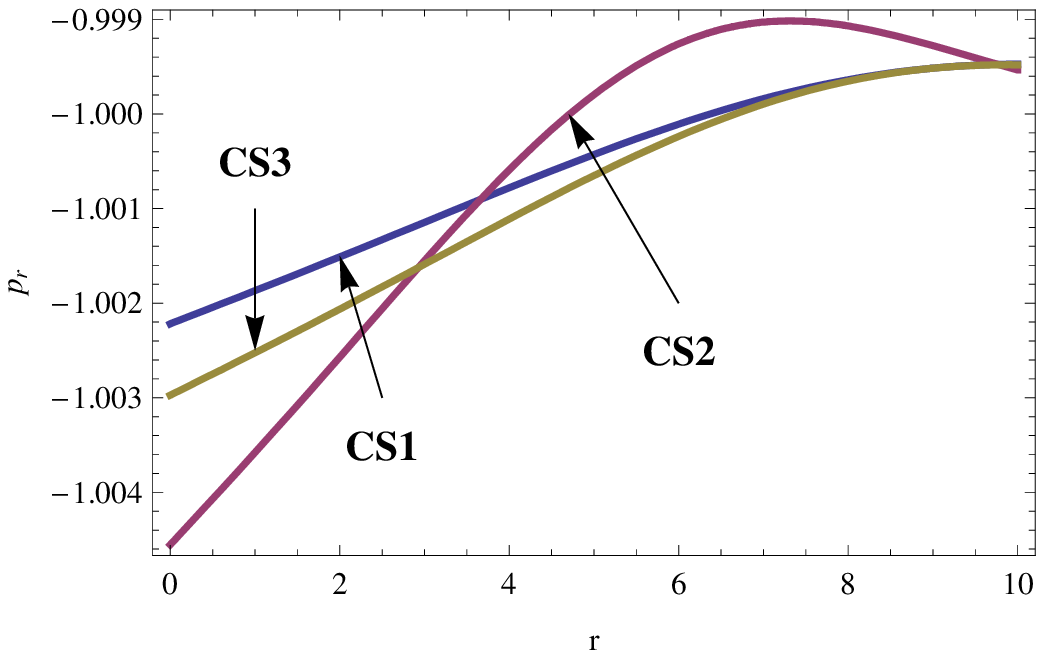}~~
\includegraphics[height=1.35in]{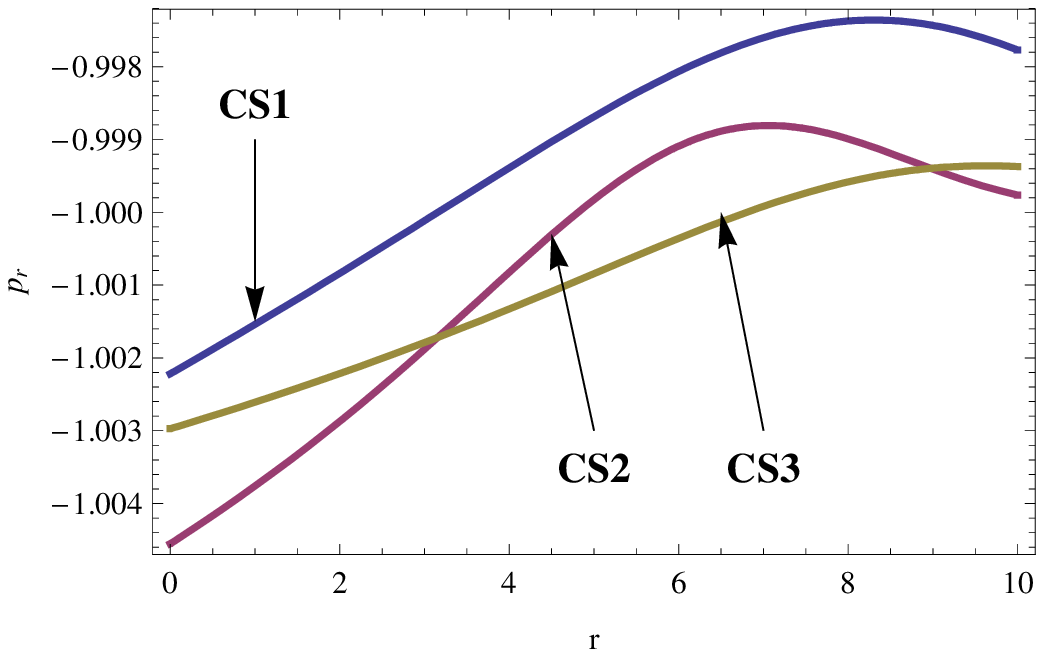}~~
\includegraphics[height=1.35in]{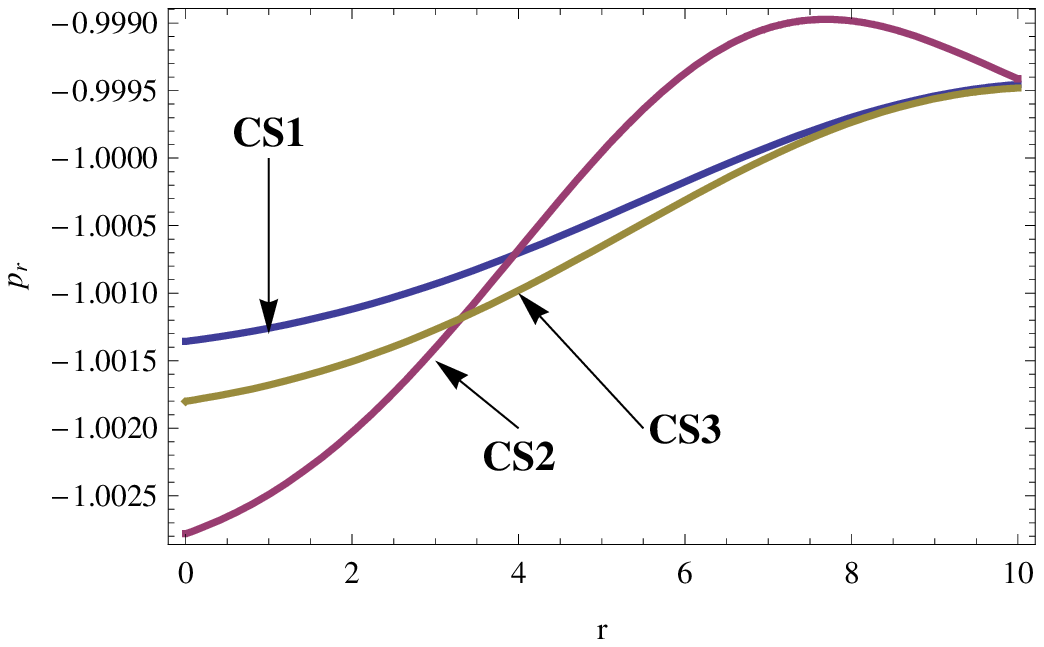}\\
\vspace{2 mm}
\textbf{Fig.2} Variations of radial pressure \textbf{$p_{r}$ ($MeV/fm^{3}$)} versus $r$ ($km$) with the numerical values of $A$, $B$ and $C$ from Table 2, Table 3 and Table 4 respectively.\\
\vspace{6 mm}

\includegraphics[height=1.35in]{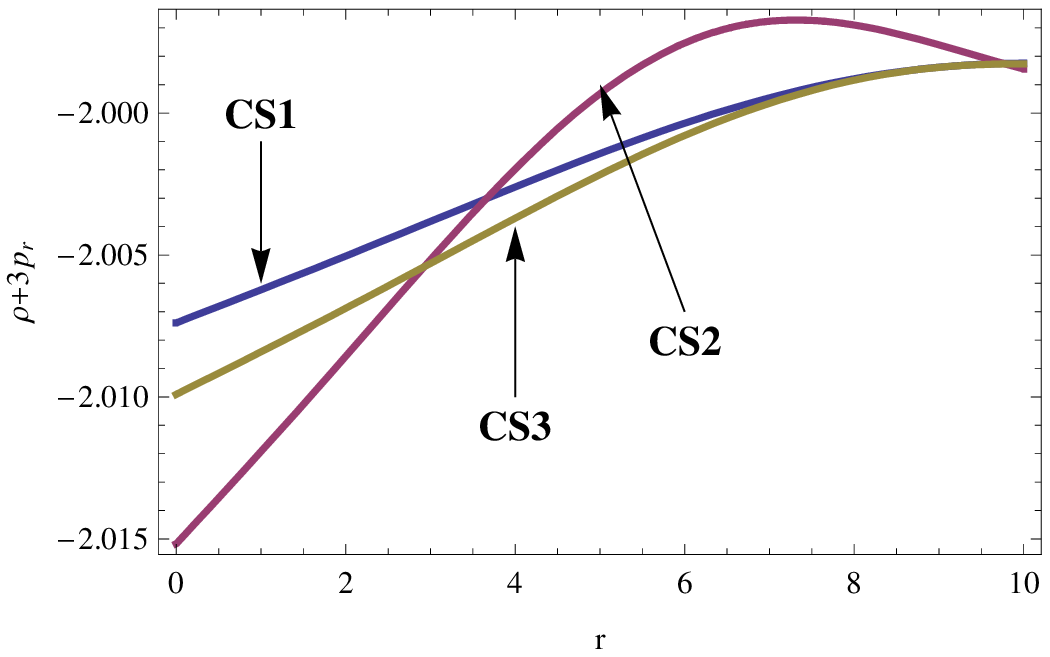}~~
\includegraphics[height=1.35in]{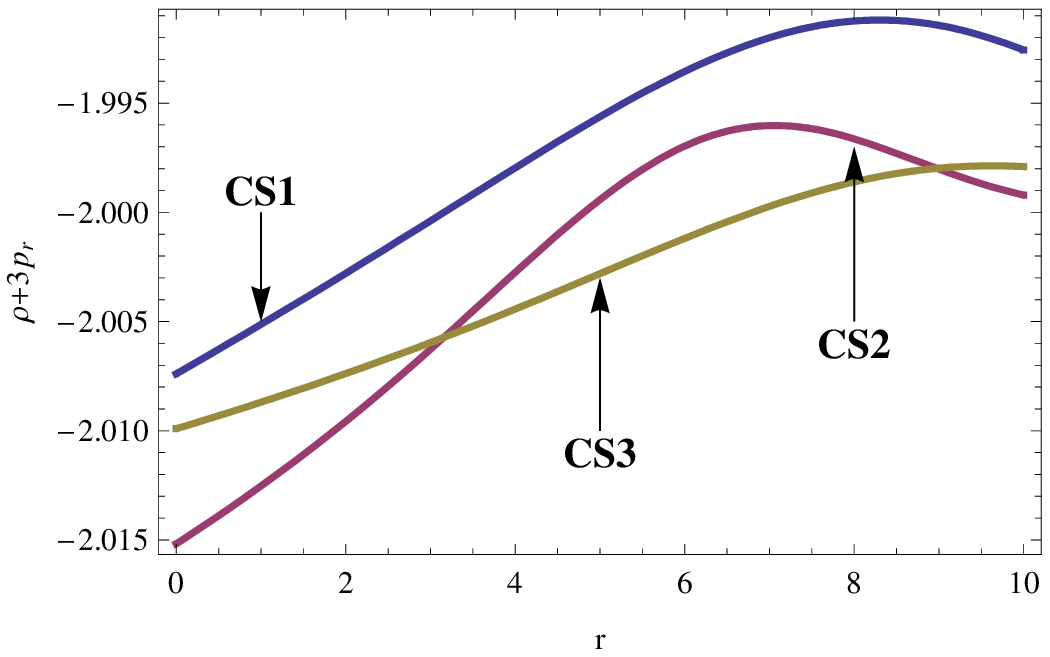}~~
\includegraphics[height=1.35in]{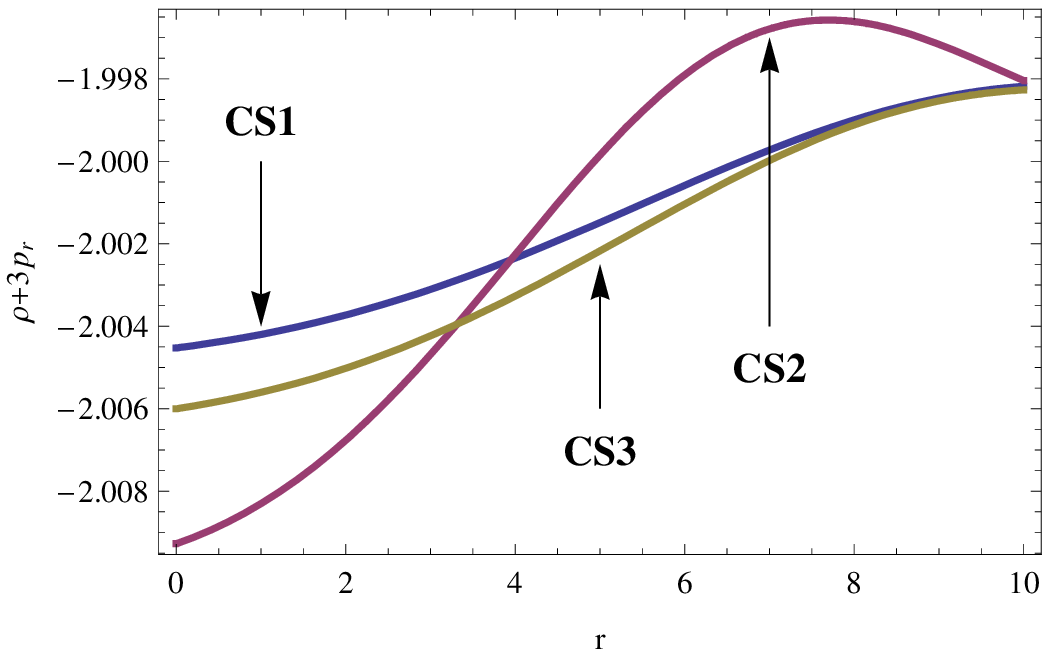}\\
\vspace{2 mm}
\textbf{Fig.3} Variations of \textbf{$\rho+3p_{r}$ ($MeV/fm^{3}$)} versus $r$ ($km$) with the numerical values of $A$, $B$ and $C$ from Table 2, Table 3 and Table 4 respectively.\\
\vspace{6 mm}

\includegraphics[height=1.35in]{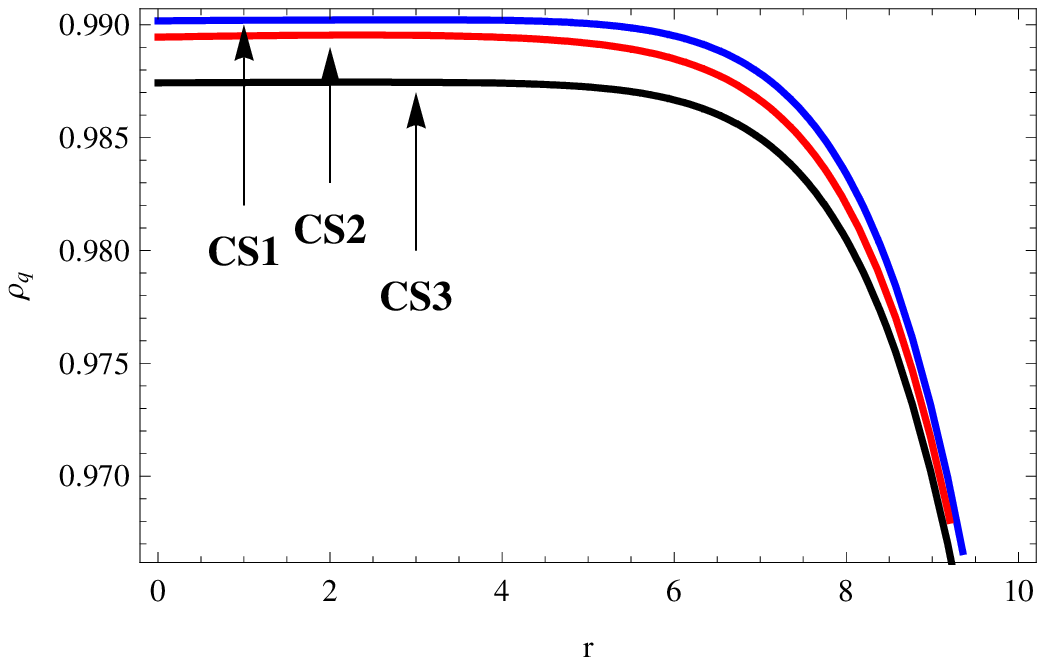}~~
\includegraphics[height=1.35in]{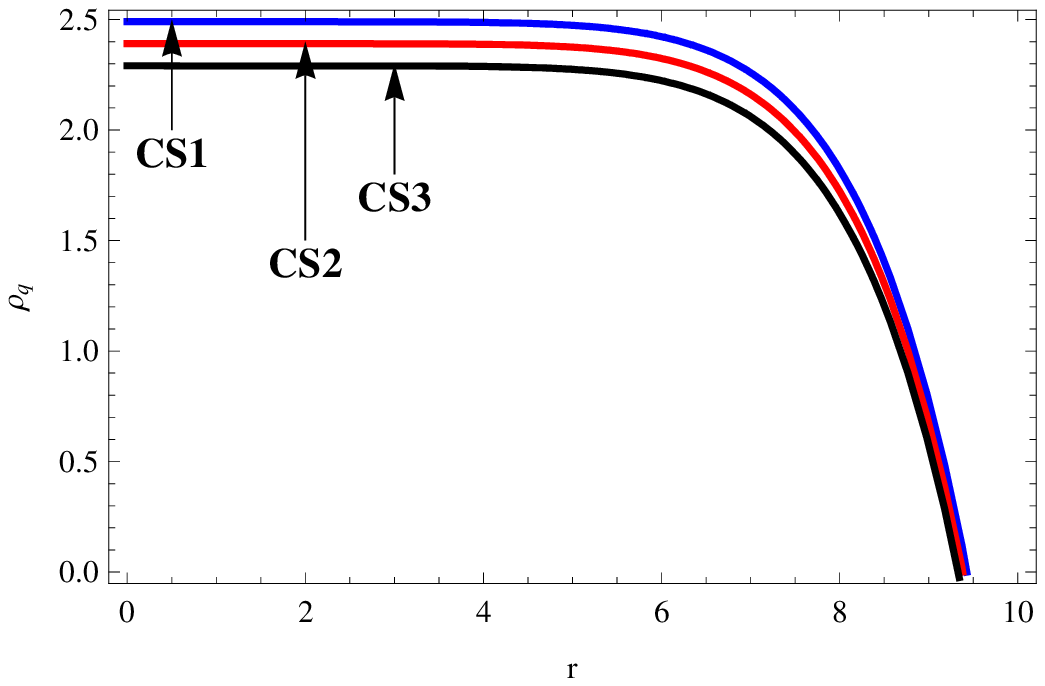}~~
\includegraphics[height=1.35in]{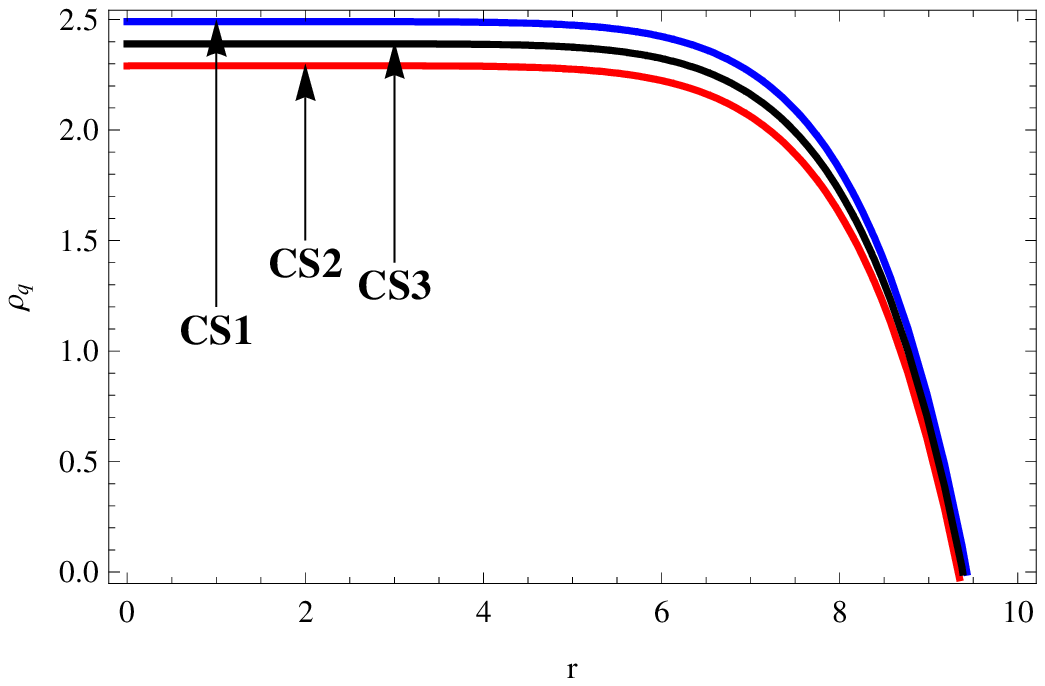}\\
\vspace{2 mm}
\textbf{Fig.4} Variations of \textbf{$\rho_{q}$ ($MeV/fm^{3}$)} versus $r$ ($km$) with the numerical values of $A$, $B$ and $C$ from Table 2, Table 3 and Table 4 respectively.\\
\vspace{6 mm}
\end{figure}

\begin{figure}

\includegraphics[height=1.35in]{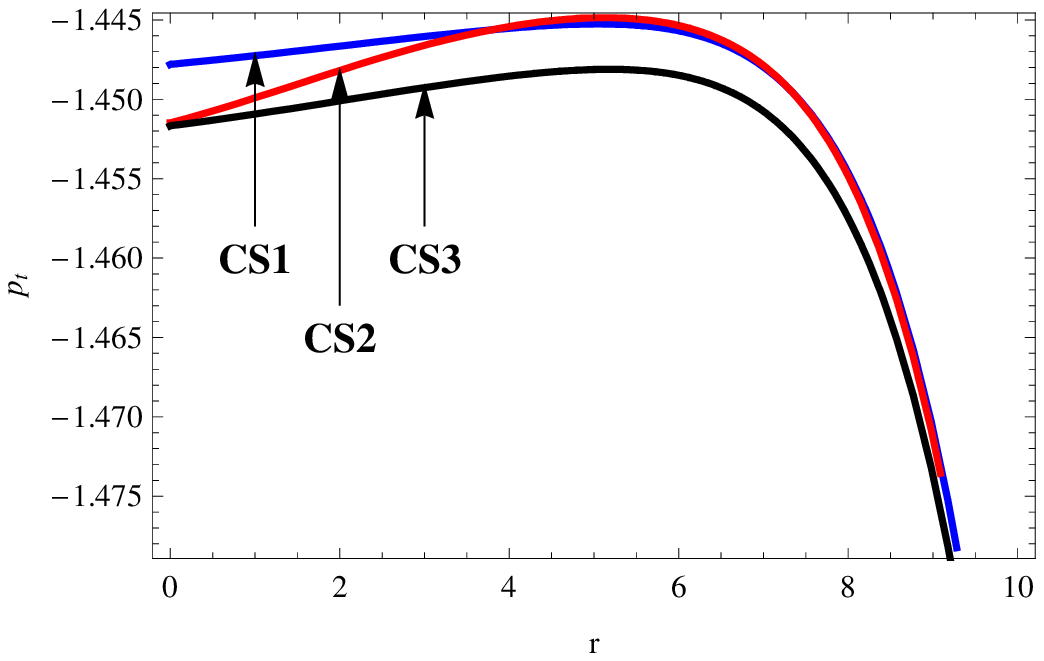}~~
\includegraphics[height=1.35in]{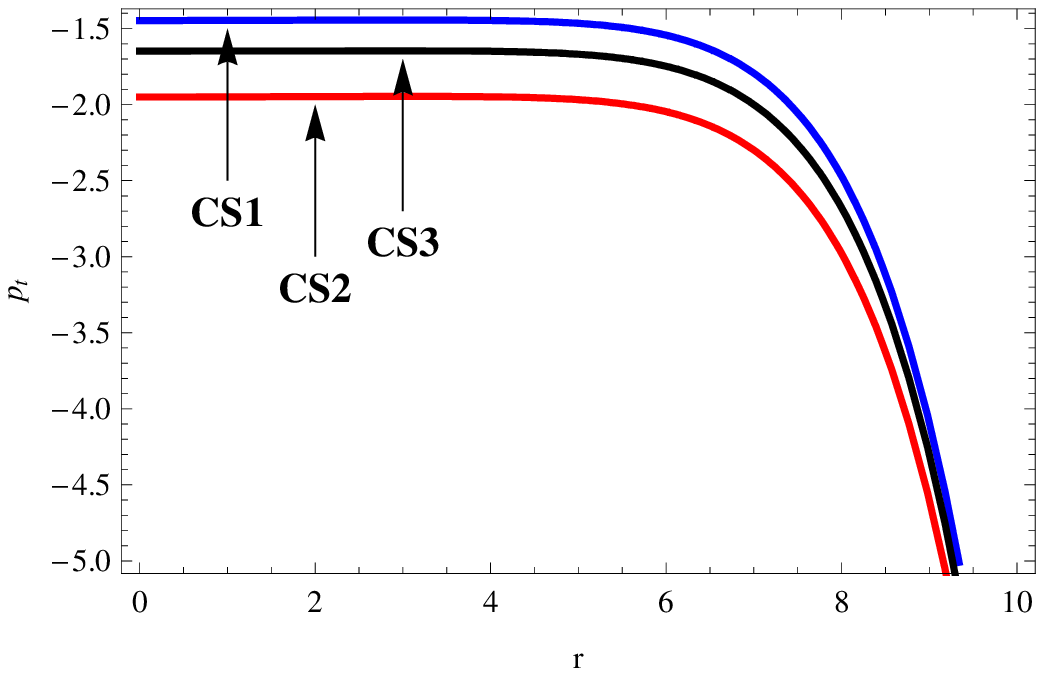}~~
\includegraphics[height=1.35in]{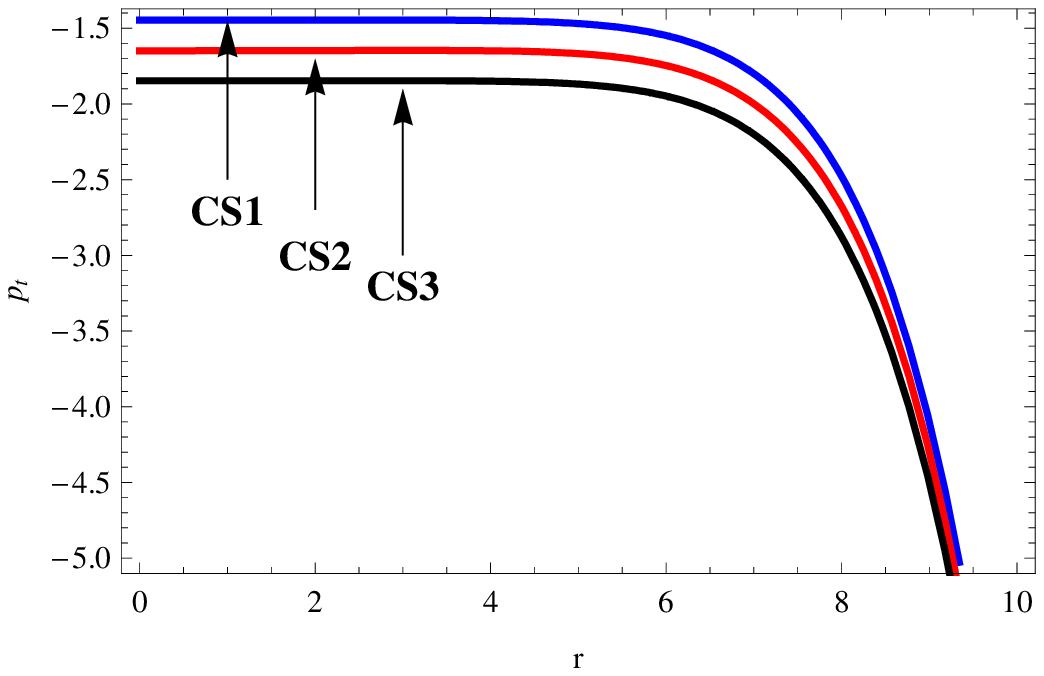}\\
\vspace{2 mm}
\textbf{Fig.5} Variations of transversal pressure \textbf{$p_{t}$ ($MeV/fm^{3}$)} versus $r$ ($km$) with the numerical values of $A$, $B$ and $C$ from Table 2, Table 3 and Table 4 respectively.\\
\vspace{6 mm}

\end{figure}

Now we have drawn different figures for physical quantities
$\rho$, $p_{r}$, $\rho+3p_{r}$, $\rho_{q}$ and $p_{t}$ using
equations (29)-(32) with the values of different values of $A$,
$B$ and $C$ taking from Tables 2-4. Figs.1-3 ensure that our
proposed model is a candidate of dark energy as $\rho>0,
\rho+3p_{r}<0$ \textbf{i.e., $w_{r}<-\frac{1}{3}$ (See equation (33))}. Again, from Fig.3, we can conclude that our model
corresponds to quintessence dark energy model, not phantom dark
energy candidate
due to $\rho+3p_{r}<0$ \textbf{i.e., $w_{r}<-\frac{1}{3}$} and $\rho+p_{r}>0$ \textbf{i.e., $w_{r}>-1$} (See Figs.10-12). We have taken quintessence field in transverse direction of anisotropic compact star in $f(T)$ gravity with modified Chaplygin gas. So, Fig.4-5 indicate that for quintessence field $\rho_{q}>0$ and $p_{t}<0$.\\

\section{Physical Analysis:}

The central density $\rho_{0}$ and central radial pressure $p_{0}$ are given by
\begin{equation*}
\rho_{0}=\lim_{r\rightarrow 0}\rho=
\left\{
\begin{array}{ll}
\frac{B\beta_{1}+\sqrt{B^{2}\beta_{1}^{2}+64\pi^{2}\zeta(1+\xi)}}{8\pi(1+\xi)}~~~~~~~~~$when$~~\alpha=2,~\beta=\gamma=3,\\
\frac{C\beta_{1}+\sqrt{C^{2}\beta_{1}^{2}+64\pi^{2}\zeta(1+\xi)}}{8\pi(1+\xi)}~~~~~~~~~$when$~~\alpha=3,~\beta=2,~\gamma=4~~$or$~~\alpha=4,~\beta=2,~\gamma=3\\
\end{array}
\right.
\end{equation*}

\begin{equation*}
p_{0}=\lim_{r\rightarrow 0}p_{r}=
\left\{
\begin{array}{ll}
\frac{-32\pi^{2}\zeta(1+\xi)+B\beta_{1}\xi\{B\beta_{1}+\sqrt{B^{2}\beta_{1}^{2}+64\pi^{2}\zeta(1+\xi)}\}}{4\pi(1+\xi)\{B\beta_{1}+\sqrt{B^{2}\beta_{1}^{2}+64\pi^{2}\zeta(1+\xi)}\}}~~~~~~~~~$when$~~\alpha=2,~\beta=\gamma=3,\\
\frac{-32\pi^{2}\zeta(1+\xi)+C\beta_{1}\xi\{C\beta_{1}+\sqrt{C^{2}\beta_{1}^{2}+64\pi^{2}\zeta(1+\xi)}\}}{4\pi(1+\xi)\{C\beta_{1}+\sqrt{C^{2}\beta_{1}^{2}+64\pi^{2}\zeta(1+\xi)}\}}~~~~~~~~~$when$~~\alpha=3,~\beta=2,~\gamma=4,\\
~~~~~~~~~~~~~~~~~~~~~~~~~~~~~~~~~~~~~~~~~~~~~~~~~~~~~~~~~~~~~~~~~$or$~~\alpha=4,~\beta=2,~\gamma=3.\\
\end{array}
\right.
\end{equation*}

In the following subsections, we investigate the nature of the
anisotropic compact star as follows:

\subsection{Anisotropic Effects:}

The anisotropic force ($\Delta=\frac{2(p_{t}-p_{r})}{r}$) of our anisotropic quintessence compact object is
\begin{eqnarray*}
\Delta=\frac{1}{8\pi r}\Big[\beta_{1}e^{-Ar^{\gamma}}\Big\{\frac{1}{2}(\alpha(\alpha-1)Br^{\alpha-2}+\beta(\beta-1)Cr^{\beta-2})+\frac{1}{4}(\alpha Br^{\alpha-1}+\beta Cr^{\beta-1})(\alpha Br^{\alpha-1}+\beta Cr^{\beta-1}
\end{eqnarray*}
\begin{eqnarray*}
-\gamma Ar^{\gamma-1})+\frac{1}{2}(\alpha Br^{\alpha-2}+\beta Cr^{\beta-2}-\gamma Ar^{\gamma-2})\Big\}-\frac{\beta_{2}}{2}-\frac{q^{2}}{r^{4}}\Big]-\frac{1}{2}(3w_{q}+1)\frac{1}{8\pi r}\Big\{\frac{\beta_{1}e^{-Ar^{\gamma}}}{r^{2}}(\gamma Ar^{\gamma}-1)+\frac{\beta_{1}}{r^{2}}+\frac{\beta_{2}}{2}
\end{eqnarray*}
\begin{eqnarray*}
-\frac{q^{2}}{r^{4}}\Big\}-\frac{\beta_{1}e^{-Ar^{\gamma}}(\alpha Br^{\alpha}+\beta Cr^{\beta}+\gamma Ar^{\gamma})+\sqrt{\beta_{1}^{2}e^{-2Ar^{\gamma}}(\alpha Br^{\alpha}+\beta Cr^{\beta}+\gamma Ar^{\gamma})^{2}+256\pi^{2}r^{4}\zeta(1+\xi)}}{16\pi r^{3}}
\end{eqnarray*}
\begin{equation}\label{32}
-\frac{16\pi\zeta r(1+\xi)}{\beta_{1}e^{-Ar^{\gamma}}(\alpha Br^{\alpha}+\beta Cr^{\beta}+\gamma Ar^{\gamma})+\sqrt{\beta_{1}^{2}e^{-2Ar^{\gamma}}(\alpha Br^{\alpha}+\beta Cr^{\beta}+\gamma Ar^{\gamma})^{2}+256\pi^{2}r^{4}\zeta(1+\xi)}}.
\end{equation}\\

\begin{figure}

\includegraphics[height=1.35in]{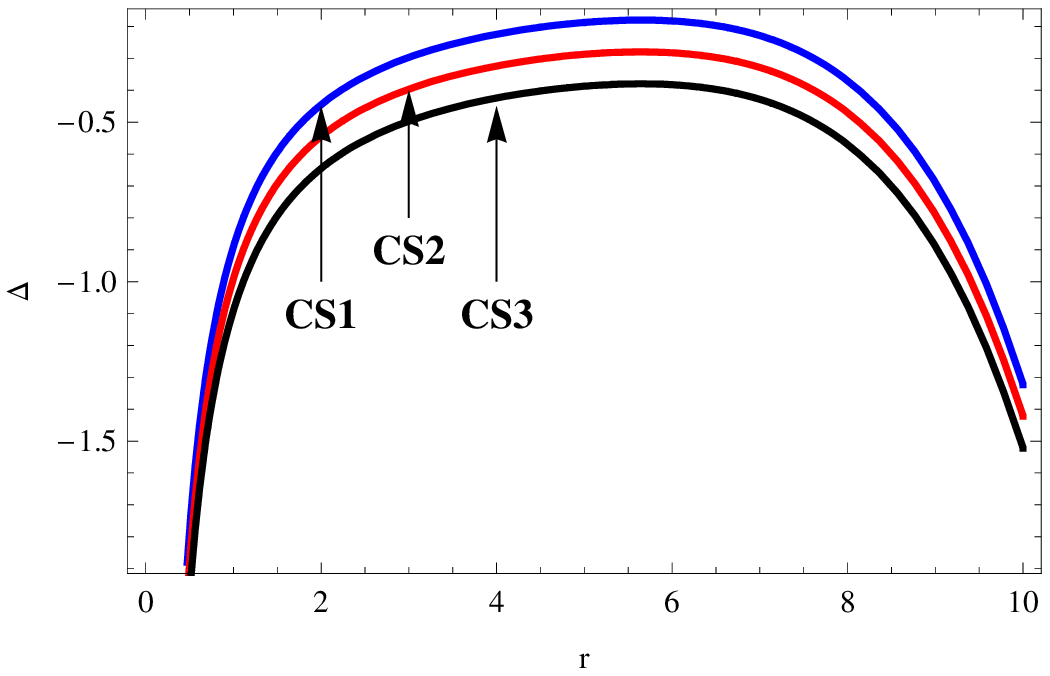}~~
\includegraphics[height=1.35in]{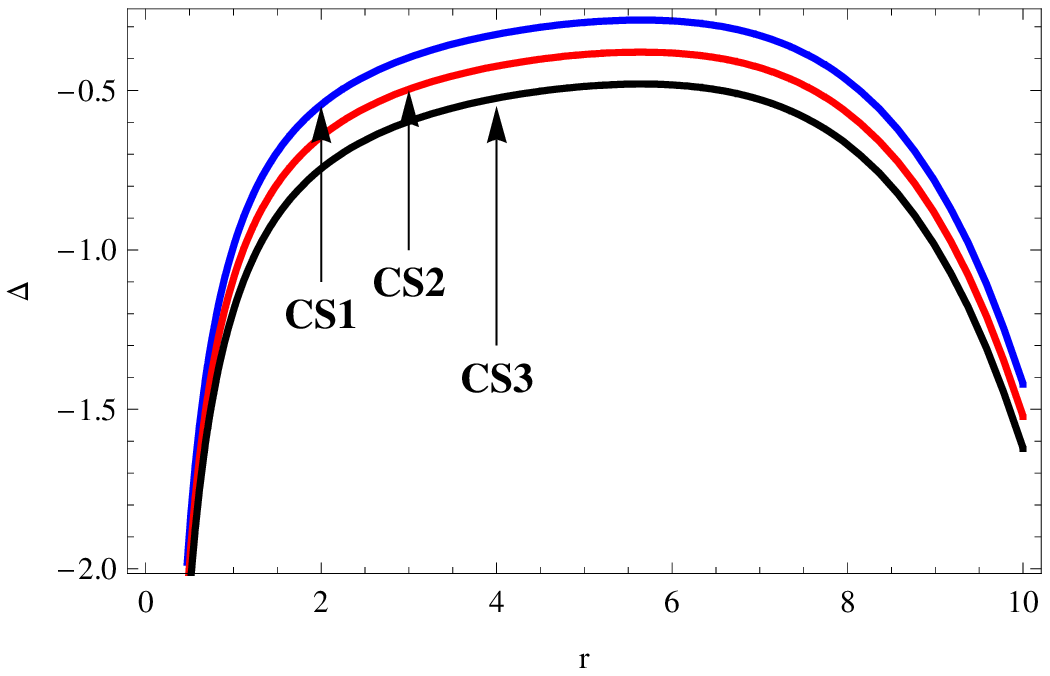}~~
\includegraphics[height=1.35in]{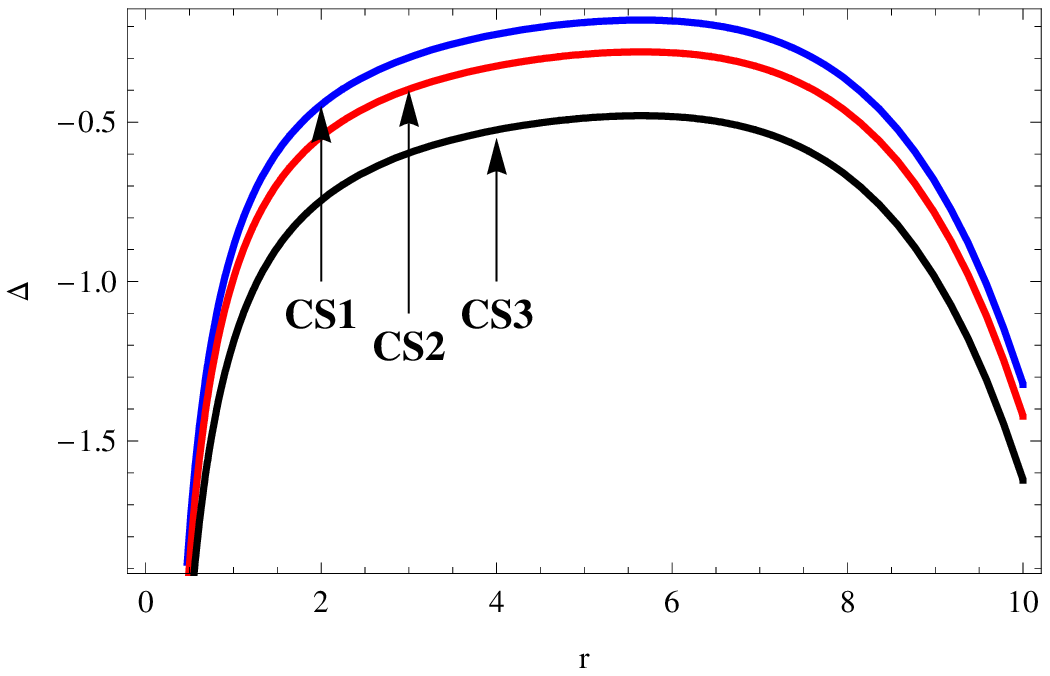}\\
\vspace{2 mm}
\textbf{Fig.6} Variations of $\Delta$ versus $r$ (km) with the numerical values of $A$, $B$ and $C$ from Table 2, Table 3 and Table 4 respectively.\\
\vspace{4 mm}

\end{figure}

From Fig.6, we have noticed that $\Delta$ is always negative i.e., $p_{t}<p_{r}$ for three compact stars (CS1, CS2, CS3) which implies that the anisotropic force is attractive like quintessence field. Since, our model consists of ordinary matter and quintessence field so the effective anisotropic force is attractive and that is why our model is very much compatible.\\

Now, we take derivatives of energy density and radial pressure with respect to radius to see where the energy density and radial pressure achieve their maximum values. With respect to three Tables 2, 3 and 4, if we make $\frac{d\rho}{dr}=0$ and $\frac{dp_{r}}{dr}=0$ then\\

\textit{From Data of Table 1:}\\
$\frac{d^{2}\rho}{dr^{2}}<0$ and $\frac{d^{2}p_{r}}{dr^{2}}<0$ at $r=10.11$ for CS1 model, at $r=7.31$ for CS2 model and at $r=9.9$ for CS3 model respectively. So, the energy density and radial pressure take maximum values at $r=10.11$ for CS1 model, at $r=7.31$ for CS2 model and at $r=9.9$ for CS3 model respectively.\\

\begin{figure}

~~~~\includegraphics[height=1.6in]{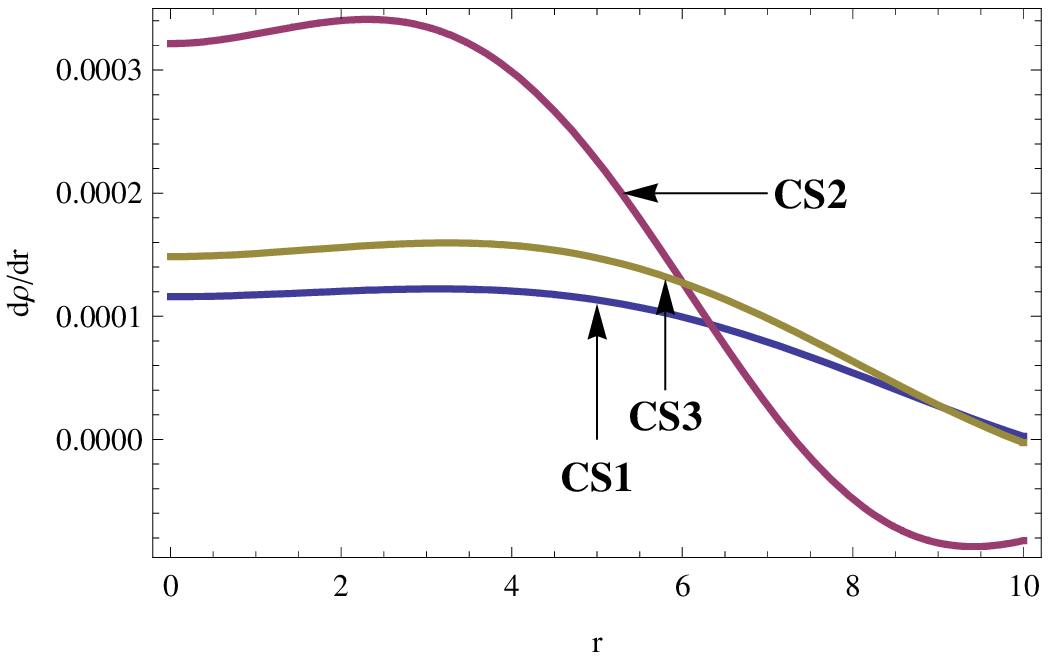}~~~~~~~~~~
\includegraphics[height=1.6in]{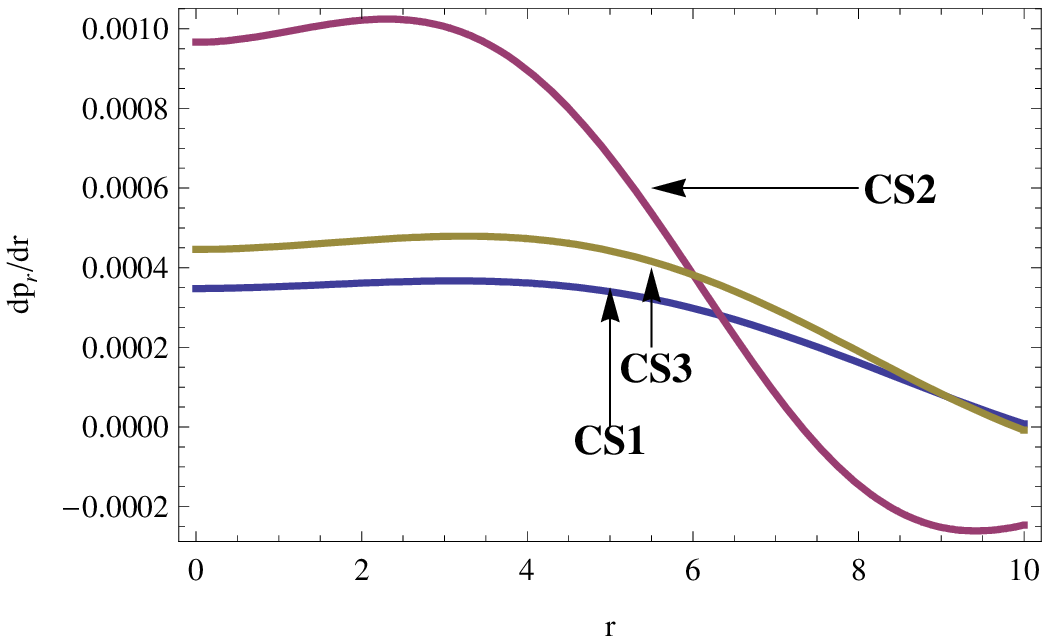}\\

\vspace{2 mm}

~\includegraphics[height=1.6in]{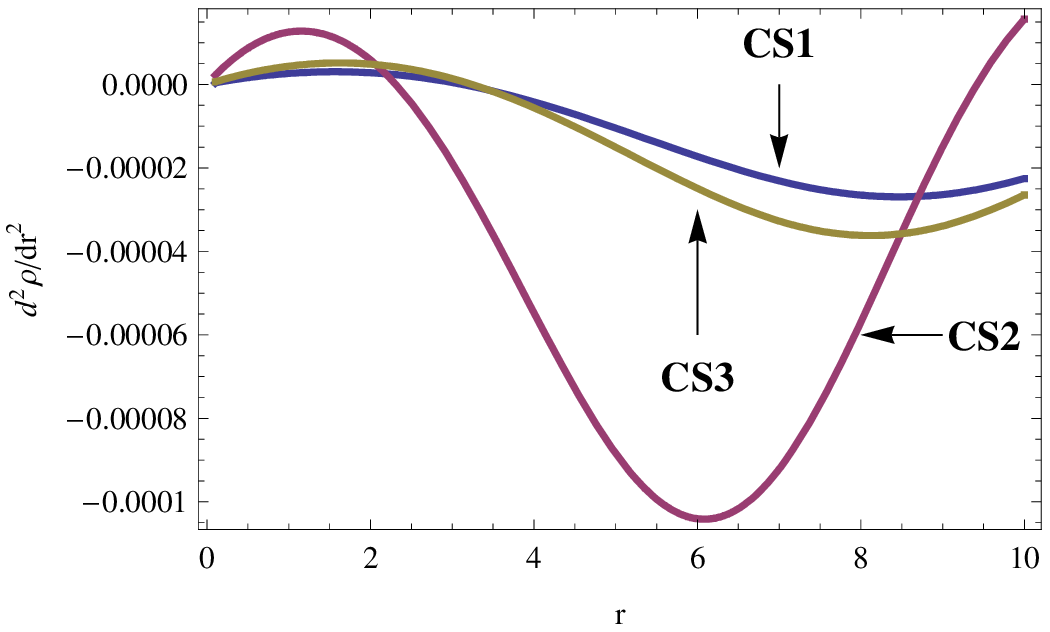}~~~~~~~~~~
\includegraphics[height=1.6in]{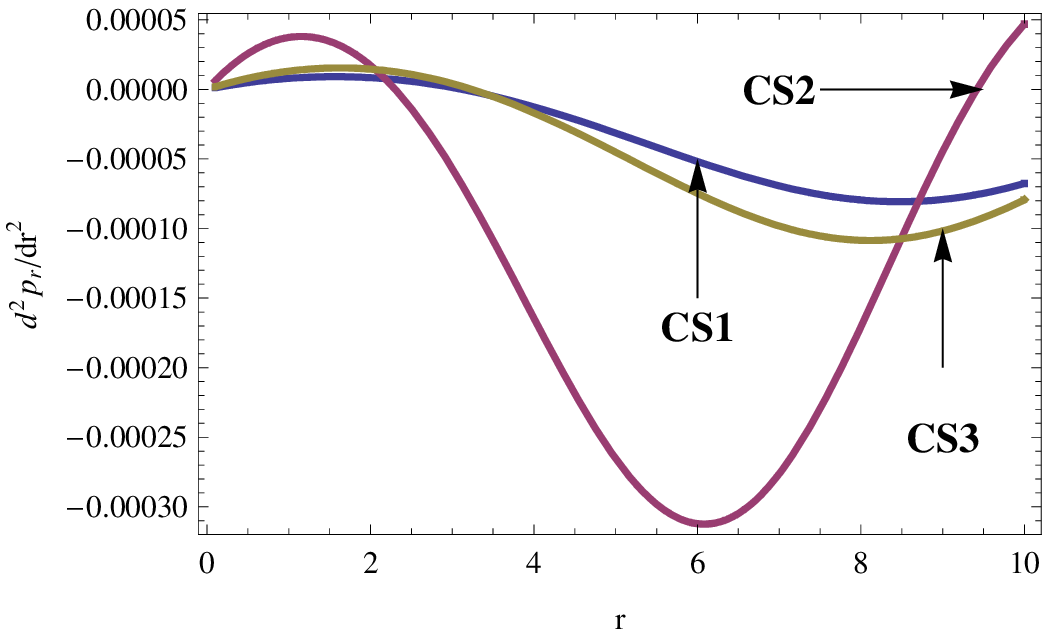}\\

\vspace{4 mm}

\textbf{Fig.7} Variations of derivatives of $\frac{d\rho}{dr}$, $\frac{dp_{r}}{dr}$, $\frac{d^{2}\rho}{dr^{2}}$ and $\frac{d^{2}p_{r}}{dr^{2}}$ with respect to $r$ (km) with the numerical values of $A$, $B$ and $C$ from Table 2.

\vspace{4 mm}
\end{figure}

\textit{From Data of Table 2:}\\
$\frac{d^{2}\rho}{dr^{2}}<0$ and $\frac{d^{2}p_{r}}{dr^{2}}<0$ at $r=8.29$ for CS1 model, at $r=7.06$ for CS2 model and at $r=9.63$ for CS3 model respectively. So, the energy density and radial pressure take maximum values at $r=8.29$ for CS1 model, at $r=7.06$ for CS2 model and at $r=9.63$ for CS3 model respectively.\\

\begin{figure}

~~~~\includegraphics[height=1.6in]{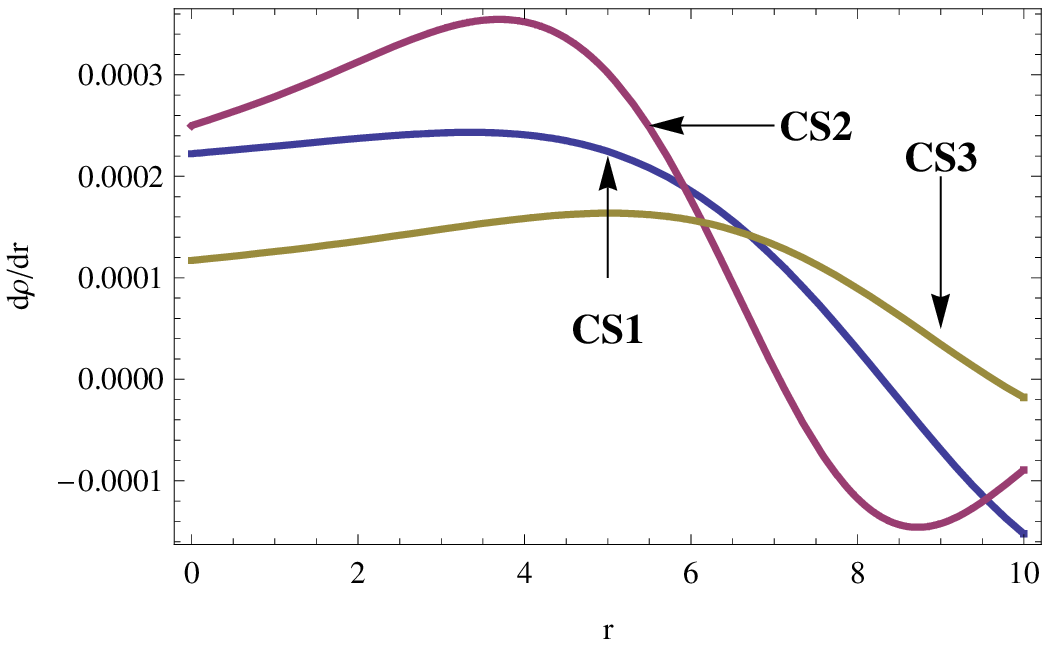}~~~~~~~~~~
\includegraphics[height=1.6in]{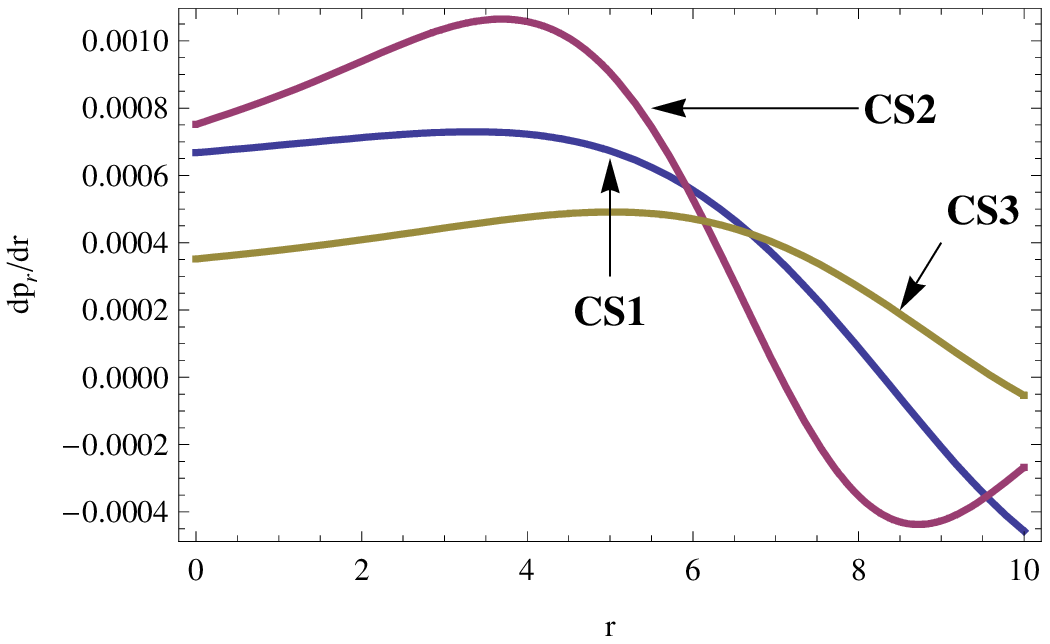}\\

\vspace{2 mm}

~\includegraphics[height=1.6in]{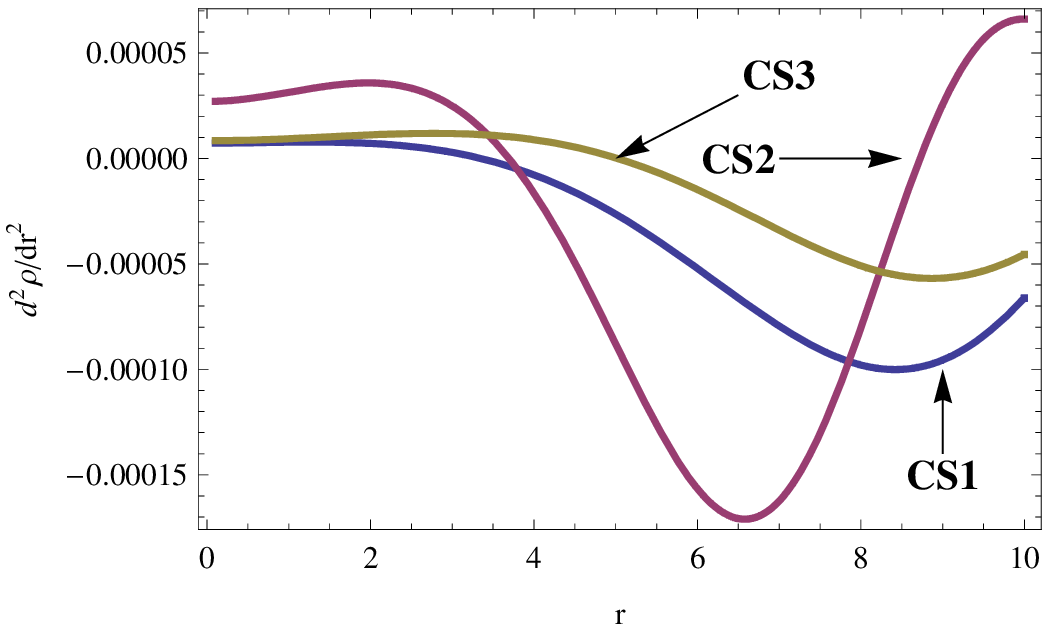}~~~~~~~~~~
\includegraphics[height=1.6in]{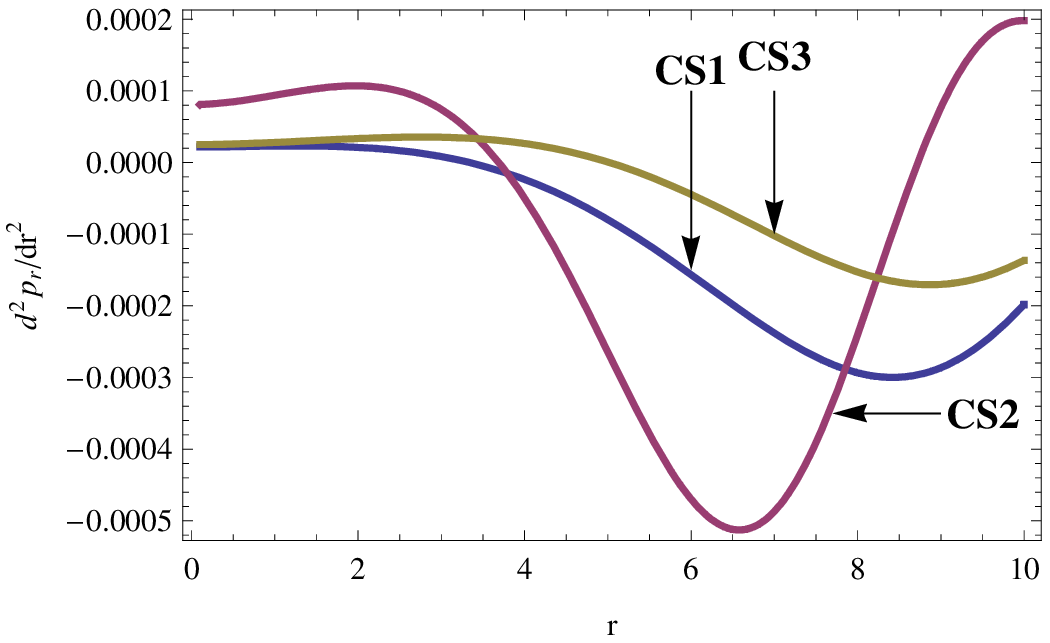}\\
\vspace{4 mm}

\textbf{Fig.8} Variations of derivatives of $\frac{d\rho}{dr}$, $\frac{dp_{r}}{dr}$, $\frac{d^{2}\rho}{dr^{2}}$ and $\frac{d^{2}p_{r}}{dr^{2}}$ with respect to $r$ (km) with the numerical values of $A$, $B$ and $C$ from Table 3.

\vspace{4 mm}
\end{figure}

\textit{From Data of Table 3:}\\
$\frac{d^{2}\rho}{dr^{2}}<0$ and $\frac{d^{2}p_{r}}{dr^{2}}<0$ at $r=10.66$ for CS1 model, at $r=7.69$ for CS2 model and at $r=10.38$ for CS3 model respectively. So, the energy density and radial pressure take maximum values at $r=10.66$ for CS1 model, at $r=7.69$ for CS2 model and at $r=10.38$ for CS3 model respectively.\\

\begin{figure}

~~~~\includegraphics[height=1.6in]{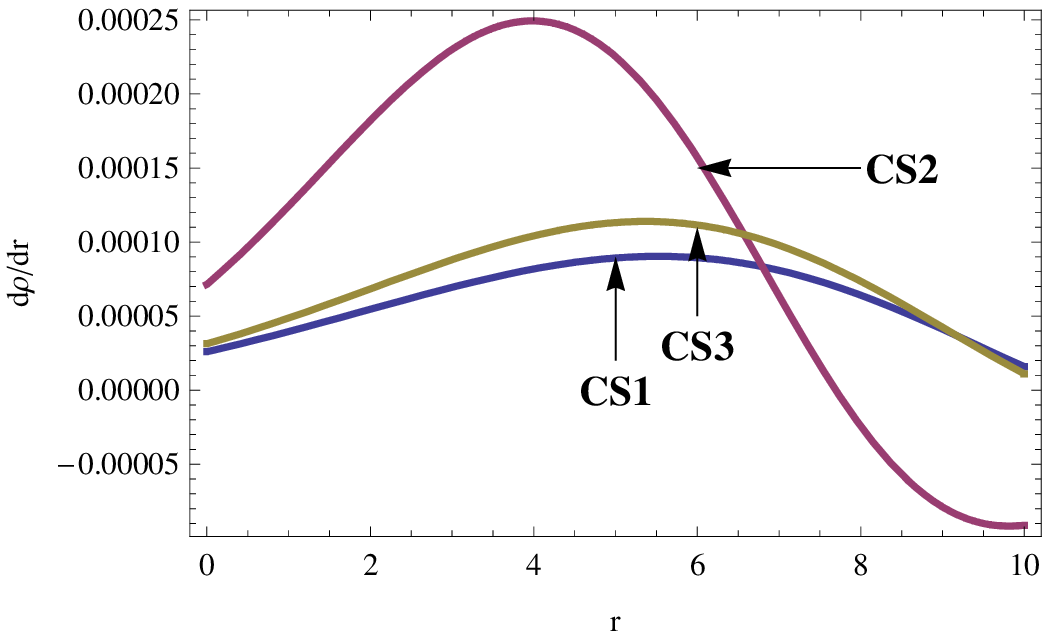}~~~~~~~~~~
\includegraphics[height=1.6in]{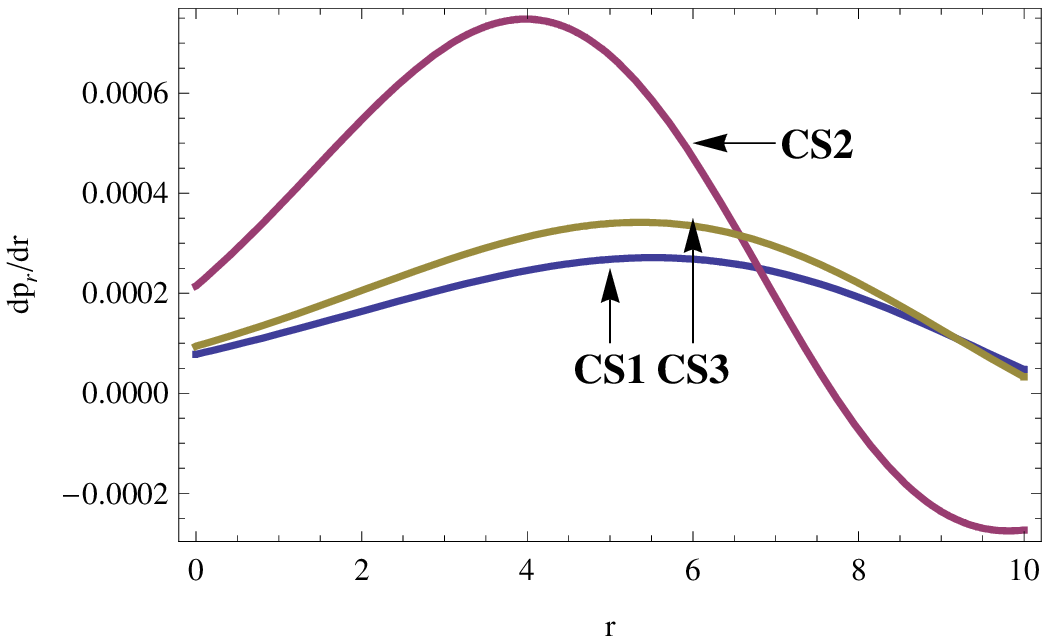}\\

\vspace{2 mm}

~\includegraphics[height=1.6in]{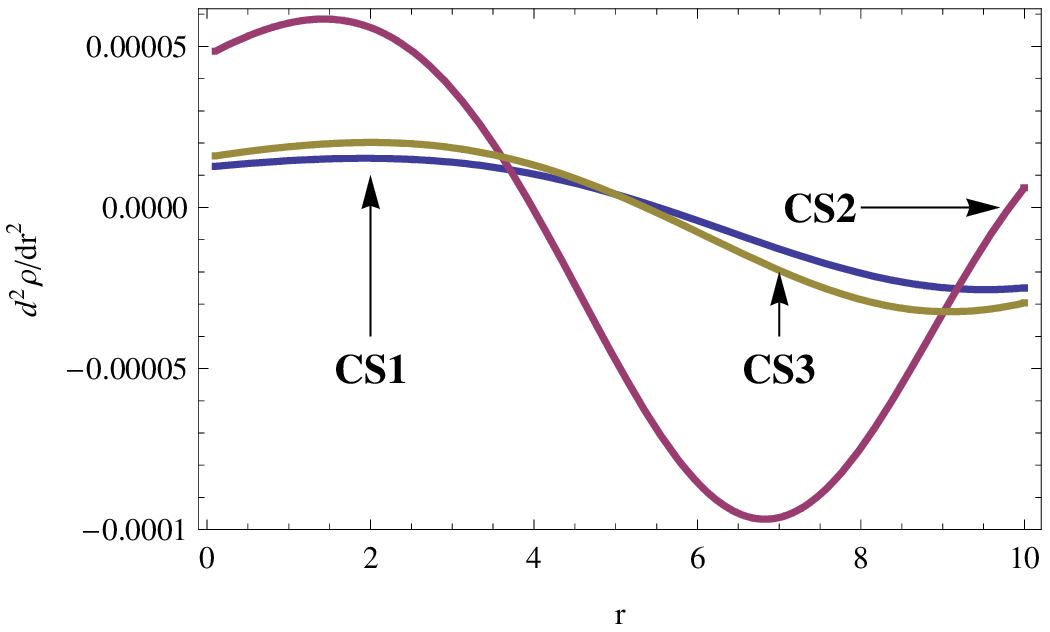}~~~~~~~~~~
\includegraphics[height=1.6in]{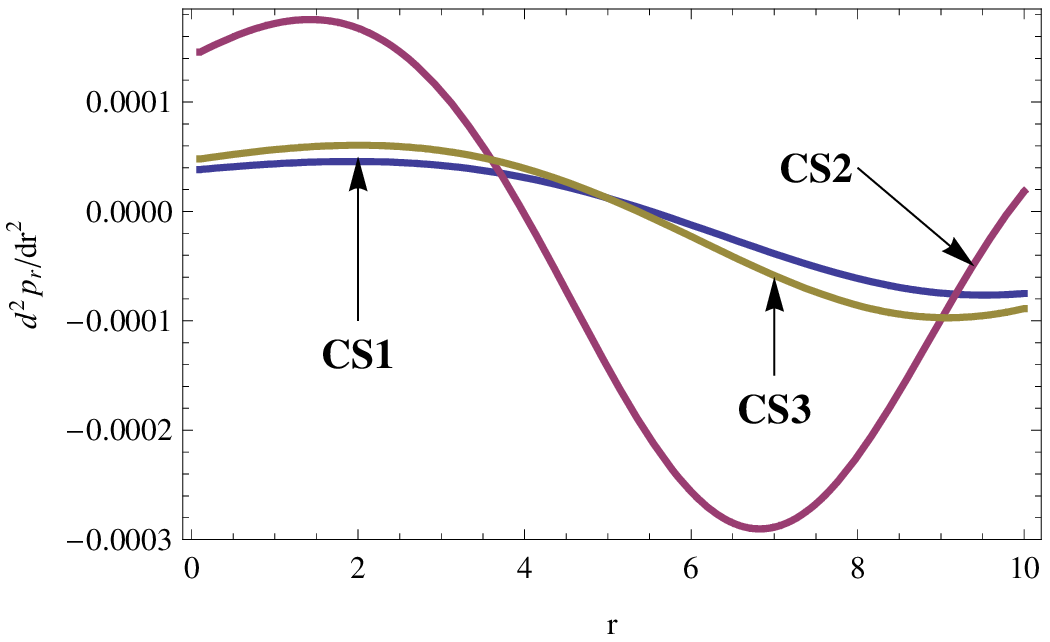}\\
\vspace{4 mm}

\textbf{Fig.9} Variations of derivatives of $\frac{d\rho}{dr}$, $\frac{dp_{r}}{dr}$, $\frac{d^{2}\rho}{dr^{2}}$ and $\frac{d^{2}p_{r}}{dr^{2}}$ with respect to $r$ (km) with the numerical values of $A$, $B$ and $C$ from Table 4.

\vspace{4 mm}

\end{figure}

However, we conclude for Table 2: with respect to Fig.7, $\frac{d\rho}{dr}>0$ always, so the energy density is increasing and $\frac{dp_{r}}{dr}$ is positive at first then changes its sign to negative, so the radial pressure is increasing first and then decreases; Table 3: with respect to Fig.8 , $\frac{d\rho}{dr}$ and $\frac{dp_{r}}{dr}$ both are positive from beginning then turn to negative sign, so consequently, the energy density and the radial pressure both are increasing first and then decrease; Table 4: with respect to Fig.9, $\frac{d\rho}{dr}$ and $\frac{dp_{r}}{dr}$ both are positive from beginning then turn to negative sign like the previous case, so the energy density and the radial pressure both are also increasing first and then decrease in this case.\\

\subsection{Energy Conditions:}

The most crucial physical properties are energy conditions to verify the existence of realistic matter distribution in this stellar model. These energy conditions are divided into three parts: Null Energy Condition (NEC), Weak Energy Condition (WEC) and Strong Energy Condition (SEC). These are very useful in general relativity and modified gravity \cite{1BP14,AG15,AGQSJA15}. The energy conditions are:

\begin{equation*}
\begin{array}{ll}
\textbf{NEC}: \rho+\frac{E^{2}}{8\pi}\geq 0,\\
\textbf{WEC}: \rho+p_{r}\geq 0,  \rho+p_{t}+\frac{E^{2}}{4\pi}\geq
0,\\
\textbf{SEC}: \rho+p_{r}+2p_{t}+\frac{E^{2}}{4\pi}\geq 0.
\end{array}
\end{equation*}\\

\begin{figure}

~~~~\includegraphics[height=1.6in]{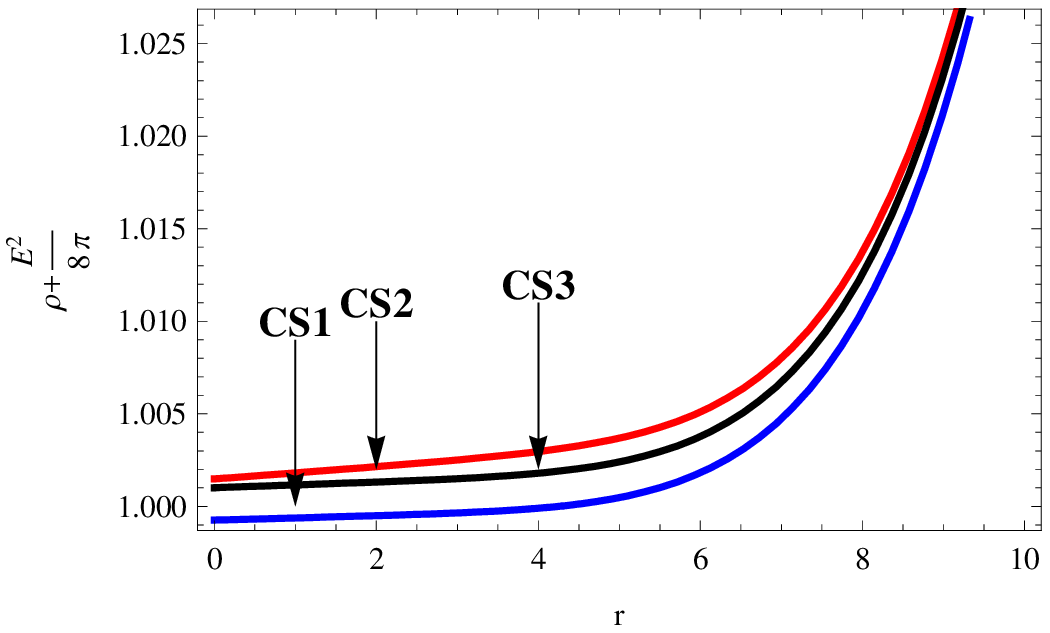}~~~~~~~~~~
\includegraphics[height=1.6in]{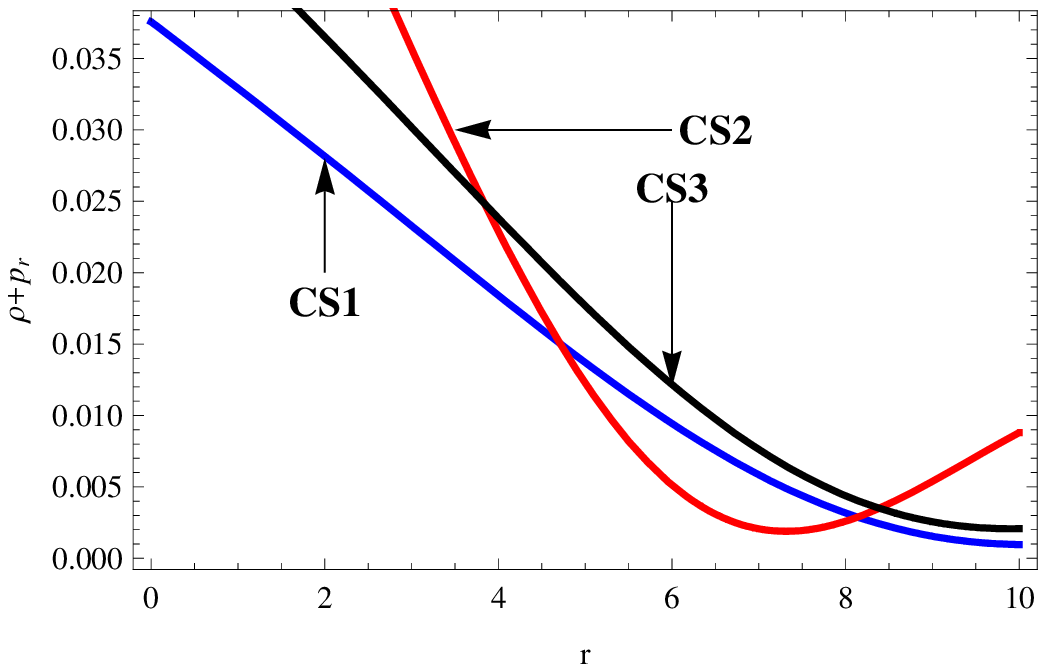}\\

\vspace{2 mm}

~\includegraphics[height=1.6in]{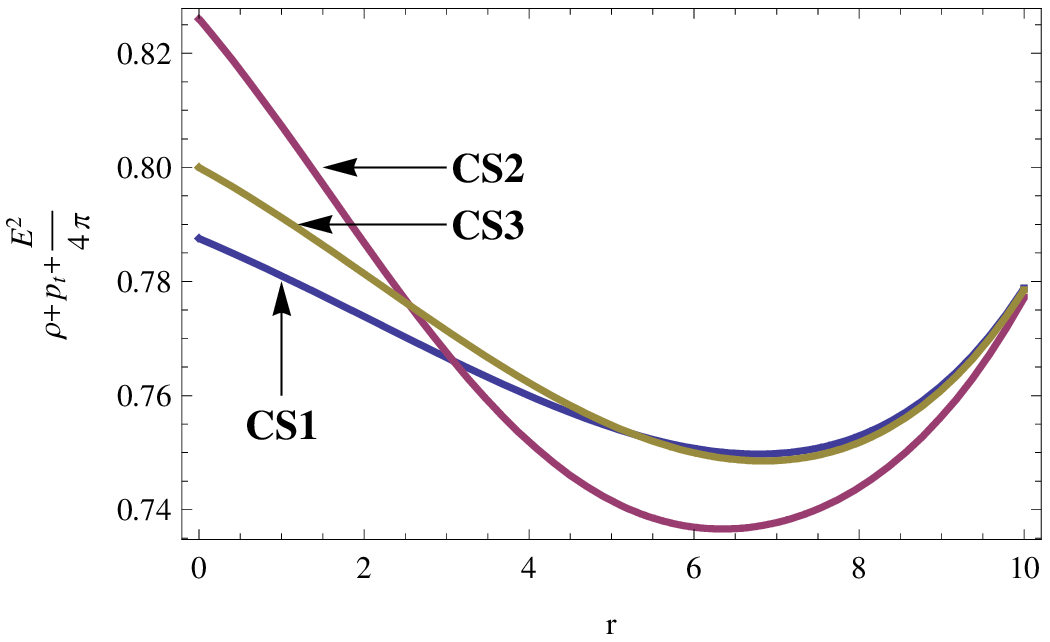}~~~~~~~~~~
\includegraphics[height=1.6in]{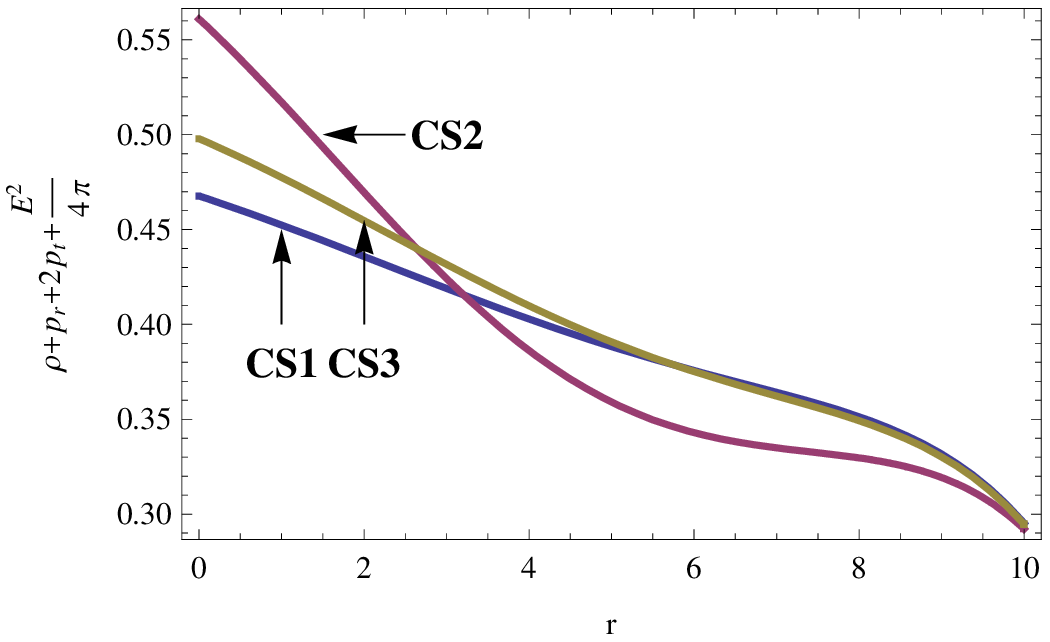}\\
\vspace{4 mm}

\textbf{Fig.10} Variations of energy conditions versus $r$ (km) with the numerical values of $A$, $B$ and $C$ from Table 2.

\vspace{4 mm}

\end{figure}

\begin{figure}

~~~~\includegraphics[height=1.6in]{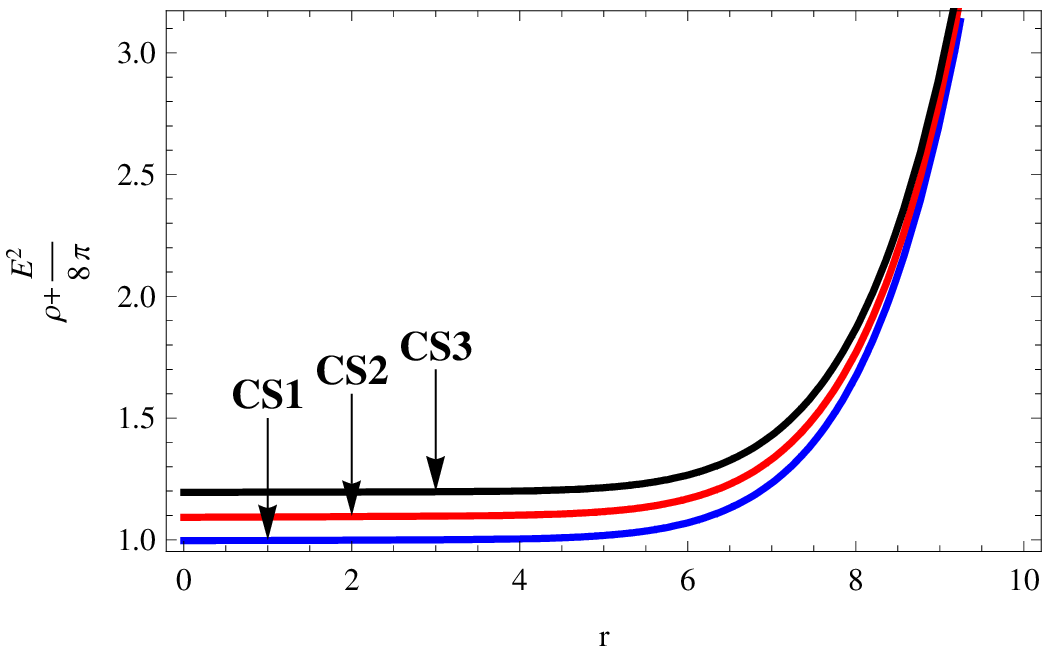}~~~~~~~~~~
\includegraphics[height=1.6in]{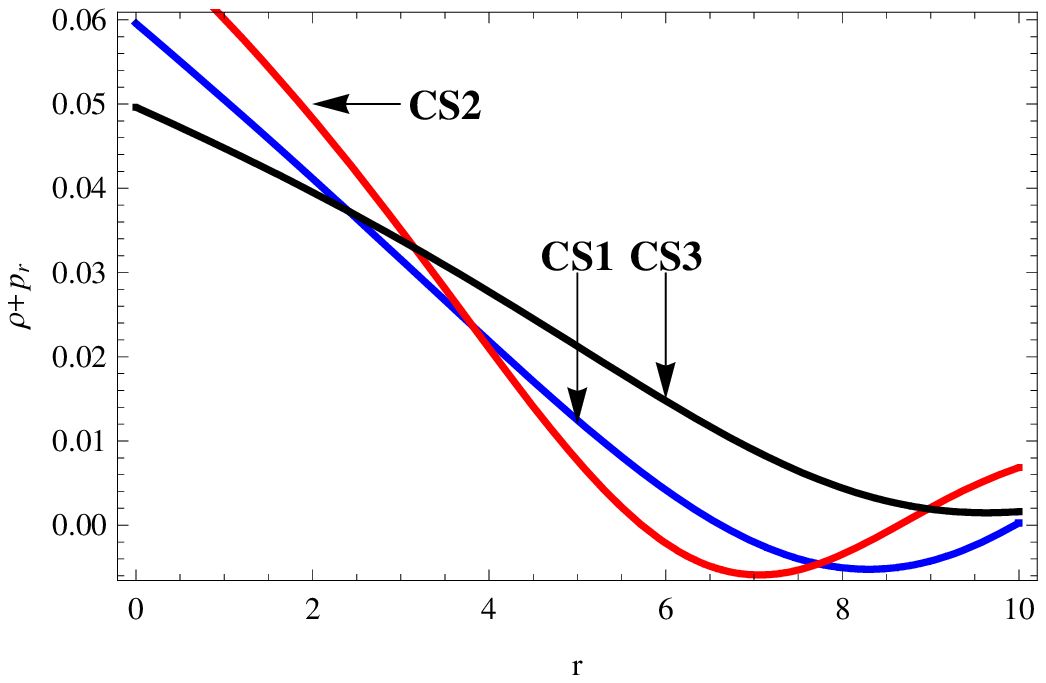}\\

\vspace{2 mm}

~~~~\includegraphics[height=1.6in]{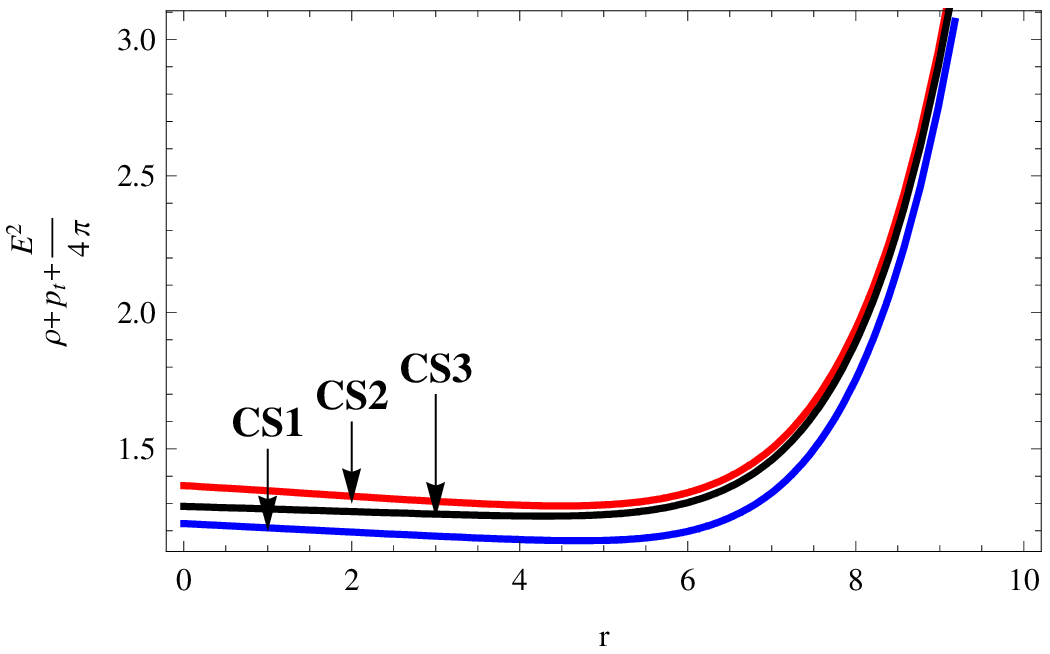}~~~~~~~~
\includegraphics[height=1.6in]{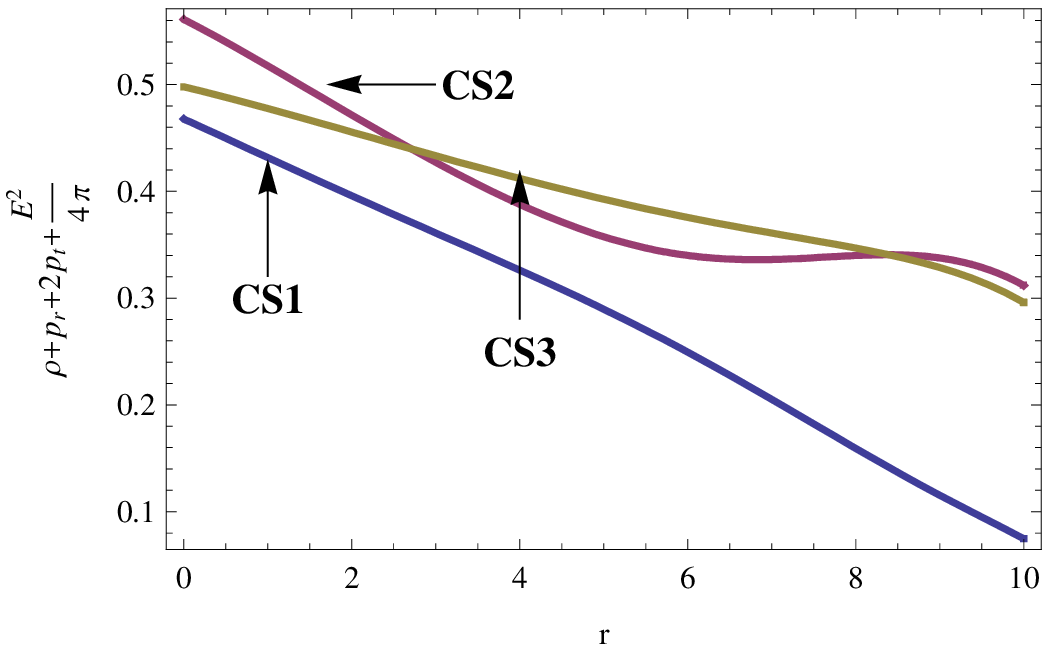}\\
\vspace{4 mm}

\textbf{Fig.11} Variations of energy conditions versus $r$ (km) with the numerical values of $A$, $B$ and $C$ from Table 3.

\vspace{4 mm}

\end{figure}

\begin{figure}

~~~~\includegraphics[height=1.6in]{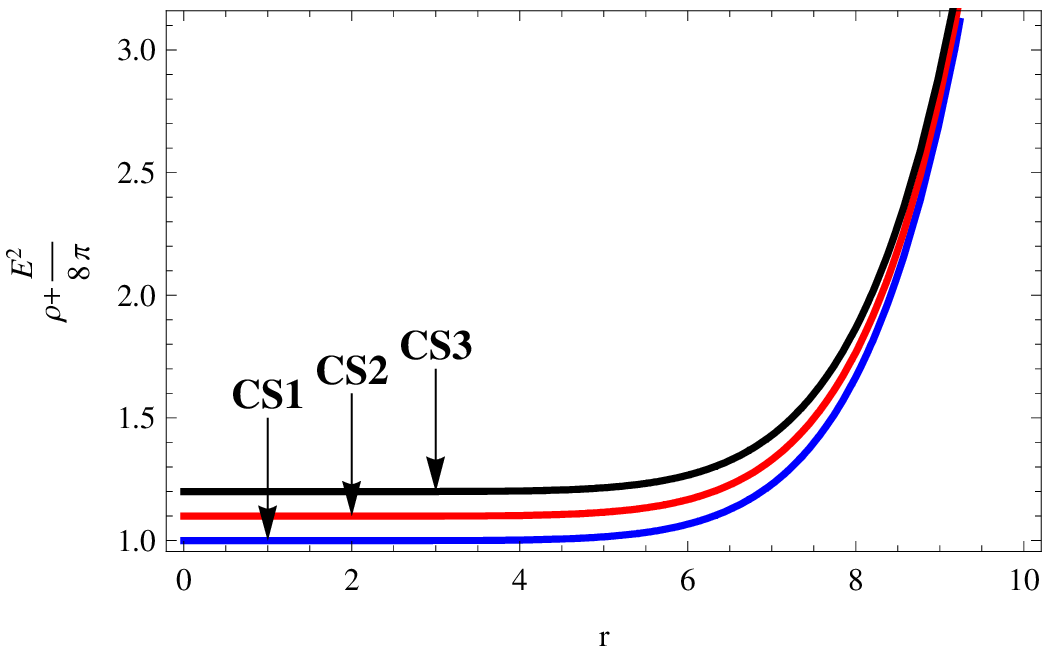}~~~~~~~~~~
\includegraphics[height=1.6in]{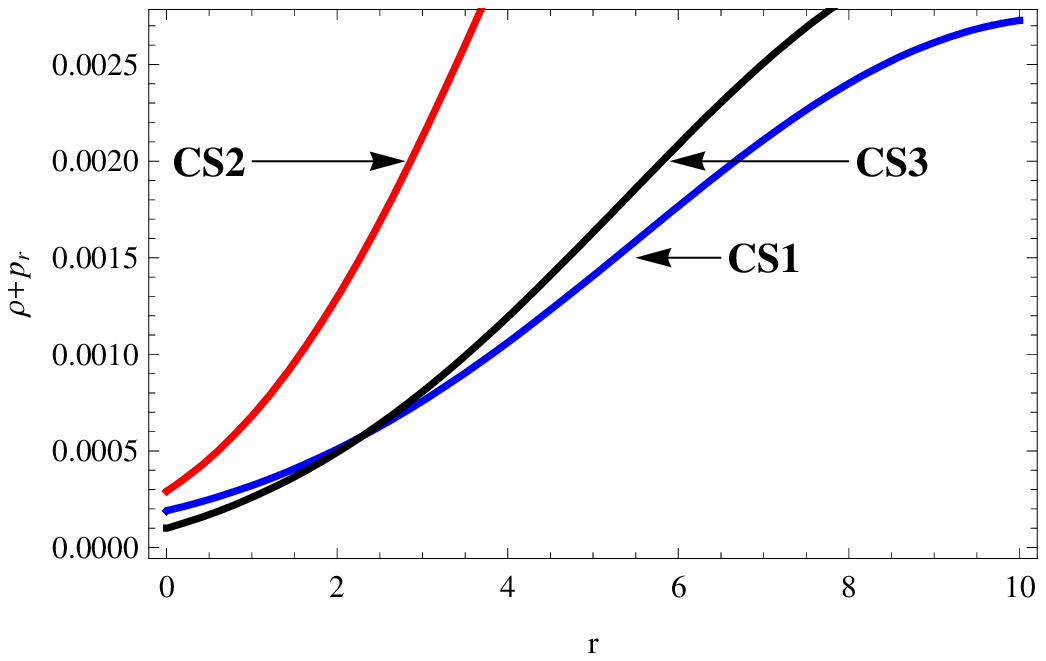}\\

\vspace{2 mm}

~~~~\includegraphics[height=1.6in]{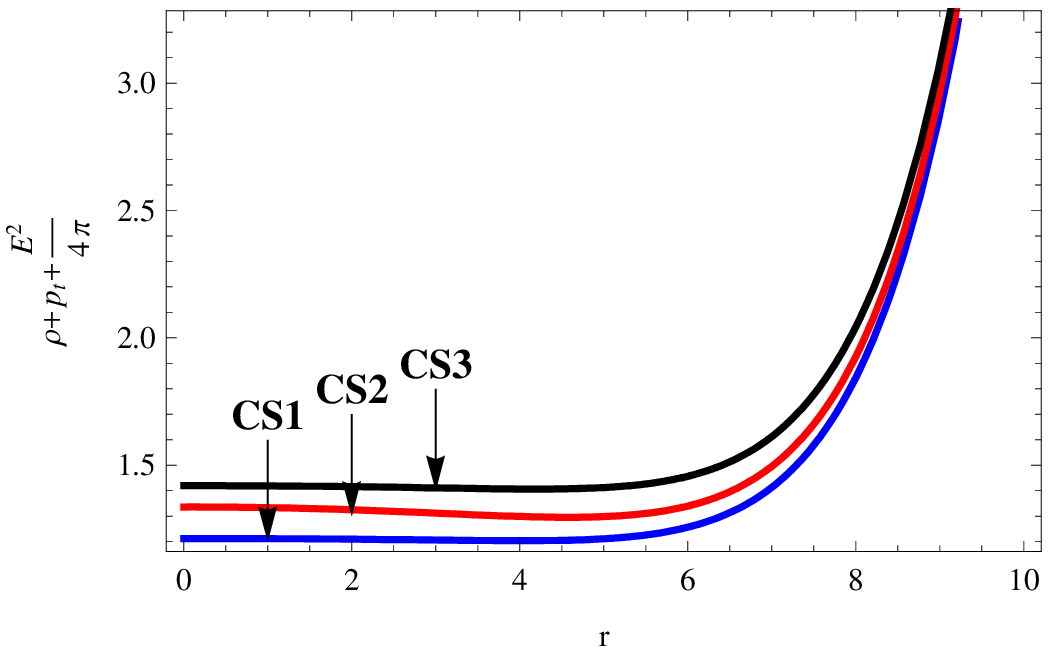}~~~~~~~~
\includegraphics[height=1.6in]{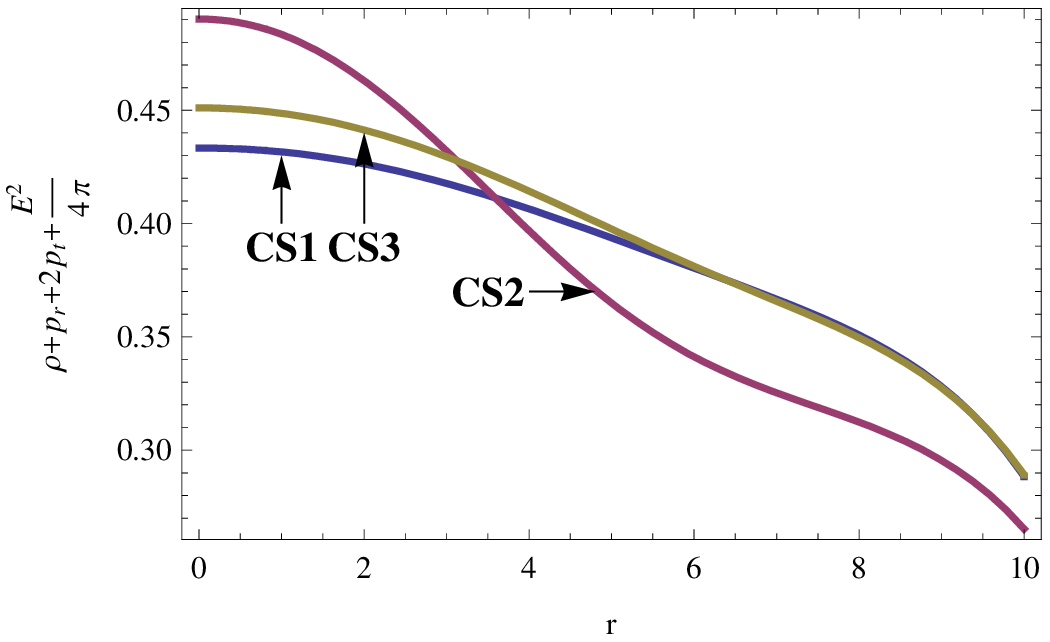}\\
\vspace{4 mm}

\textbf{Fig.12} Variations of energy conditions versus $r$ (km) with the numerical values of $A$, $B$ and $C$ from Table 4.

\vspace{4 mm}

\end{figure}

From Fig.10-12, we can conclude that our model satisfies first two conditions for arbitrary values of parameters but the last condition is satisfied only when $\beta_{2}<0$.\\

\subsection{Stability Analysis:}

To analyze stability of stellar structure against external fluctuation plays an essential key role for any physically consistent model. According to Herrera's cracking condition \cite{HL92,1BP14,AGQSJA15,AG15}, the sound speed square ($v_{s}^{2}=dp/d\rho$) must lie in the interval $[0,1]$ to be a physically stable stellar object. For our anisotropic quintessence compact star model, we have $0\leq v_{sr}^{2}\leq 1$ and $0\leq v_{st}^{2}\leq 1$ always (See Figs.13-14) where $v_{sr}^{2}$ and $v_{st}^{2}$ are sound speeds for radial and transversal components.\\

Next, we calculate the difference of two speeds of our model to examine whether transversal sound speed square is greater than the radial one or not. Actually, Herrera's cracking condition \cite{HL92} has prospected a different way to find out the potentially stable or unstable model. If the radial sound speed square is greater than the transversal sound speed square then the model is potentially stable otherwise it is potentially unstable. From Fig.15, we can decide that our model is potentially stable. Clearly, $|v_{st}^{2}-v_{sr}^{2}|\leq 1$ is verified \cite{1AH09} from Fig.16.\\

We can check stability of our proposed model by adiabatic index \cite{CS64,HHHW75,HWSKO76,BI96} which depicts the stiffness of the EoS parameter for our model. By this theory, many researchers have examined the dynamical stability against infinitesimal radial perturbation of the realistic as well as non-realistic stellar objects. For anisotropic fluid, the adiabatic index is defined as
\begin{equation*}
\Gamma=\Big(1+\frac{\rho}{p_{r}}\Big)\frac{dp_{r}}{d\rho}.
\end{equation*}
According to \cite{CS64}, the evaluated value of this adiabatic index should be greater than $4/3$. Fig.17 ensures that our proposed model satisfies this condition i.e., our anisotropic quintessence compact star model in $f(T)$ gravity with modified Chaplygin gas maintains stability range also.\\

\begin{figure}

\includegraphics[height=1.35in]{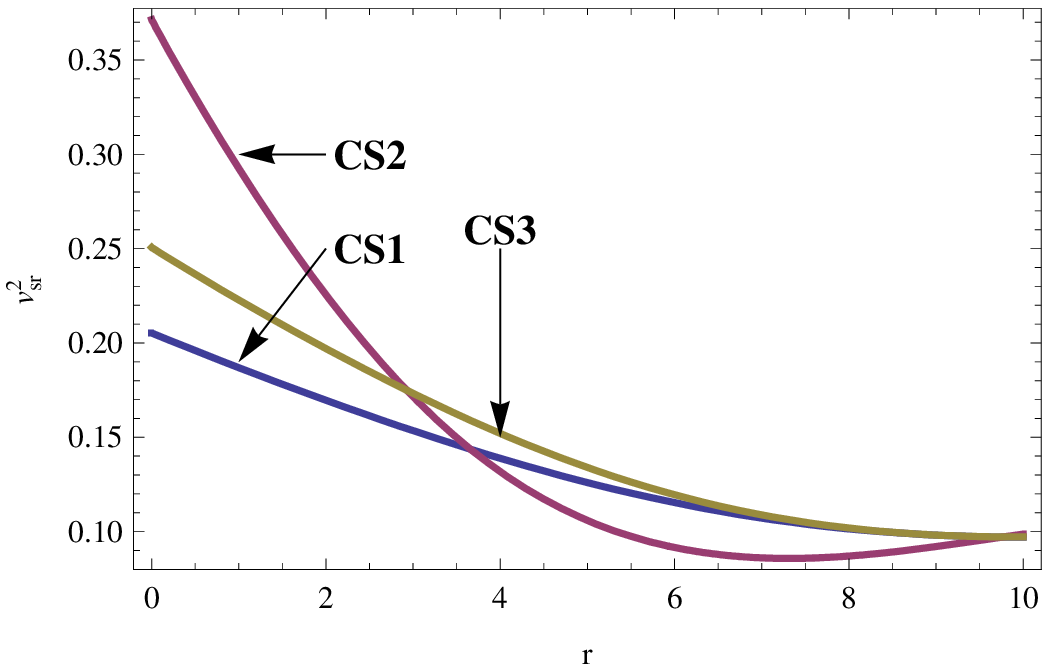}~~
\includegraphics[height=1.35in]{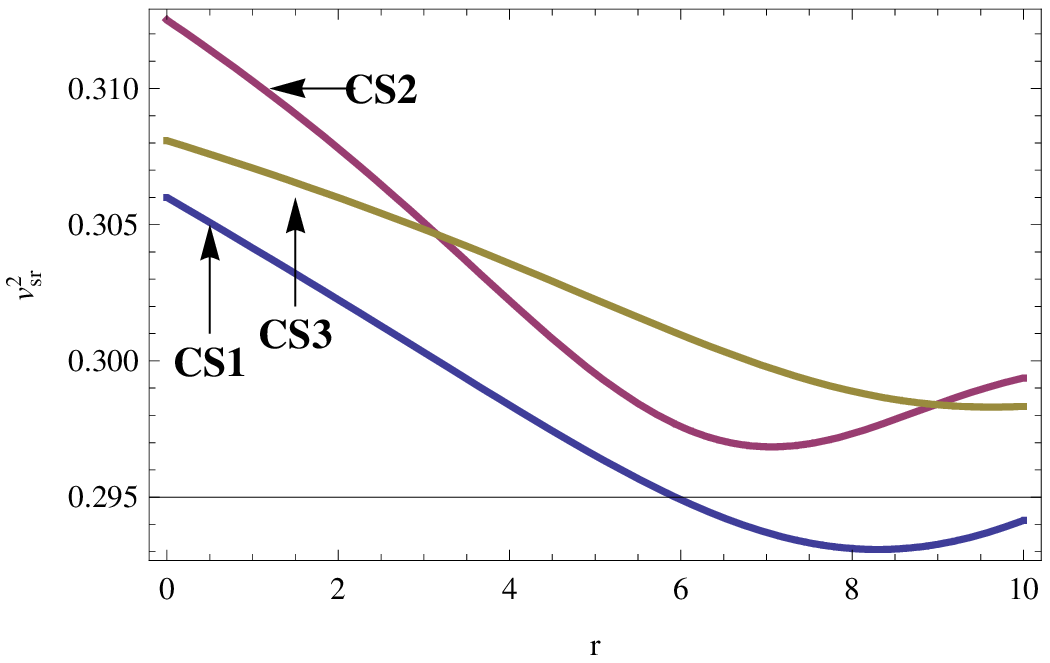}~~
\includegraphics[height=1.35in]{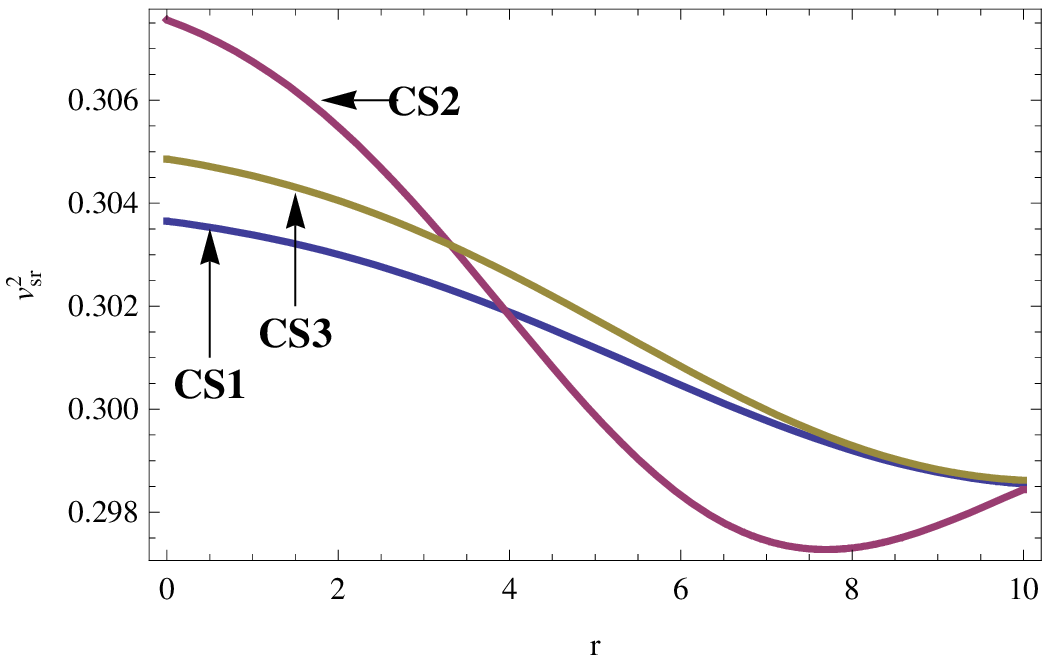}\\
\vspace{2 mm}
\textbf{Fig.13} Variations of $v_{sr}^{2}$ versus $r$ (km) with the numerical values of $A$, $B$ and $C$ from Table 2, Table 3 and Table 4 respectively.\\
\vspace{4 mm}

\end{figure}

\begin{figure}

\includegraphics[height=1.35in]{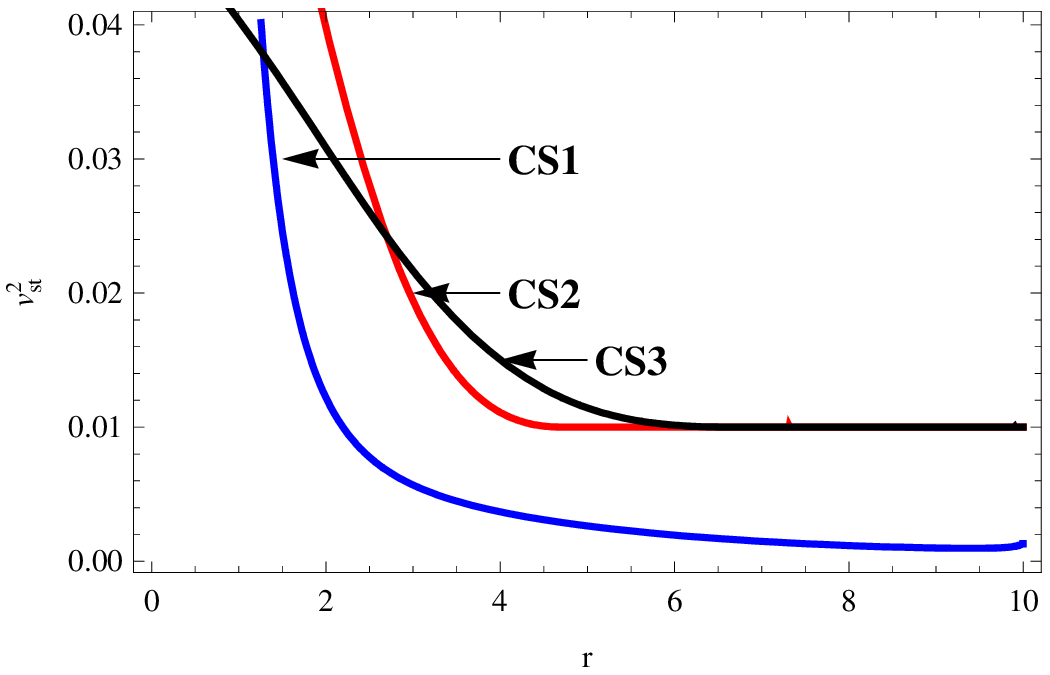}~~~~
\includegraphics[height=1.35in]{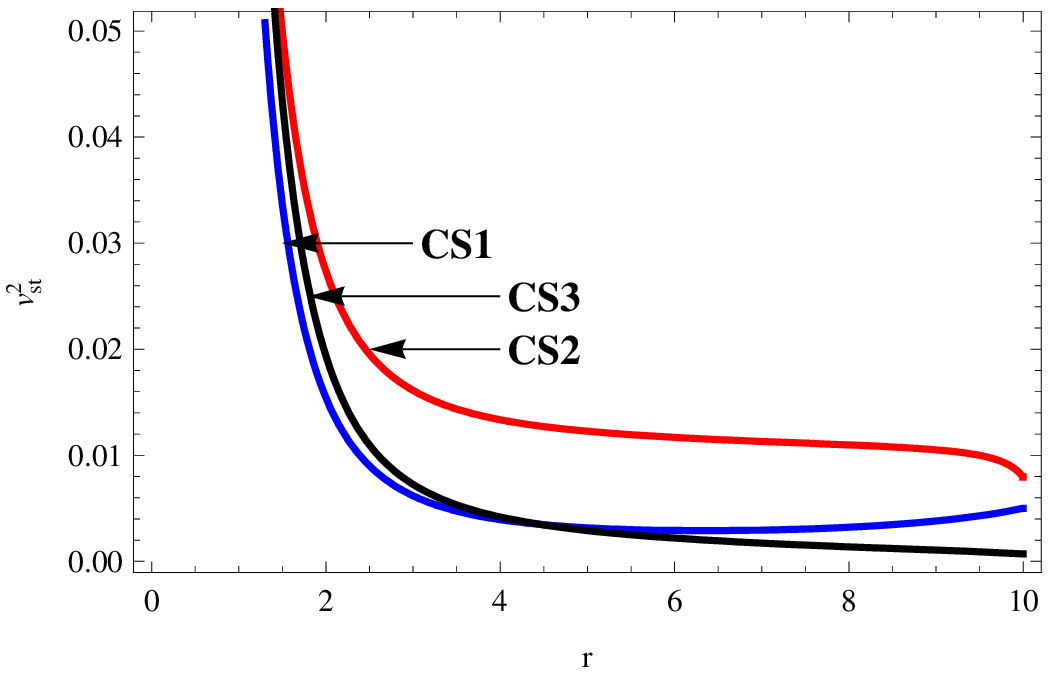}~~~~
\includegraphics[height=1.35in]{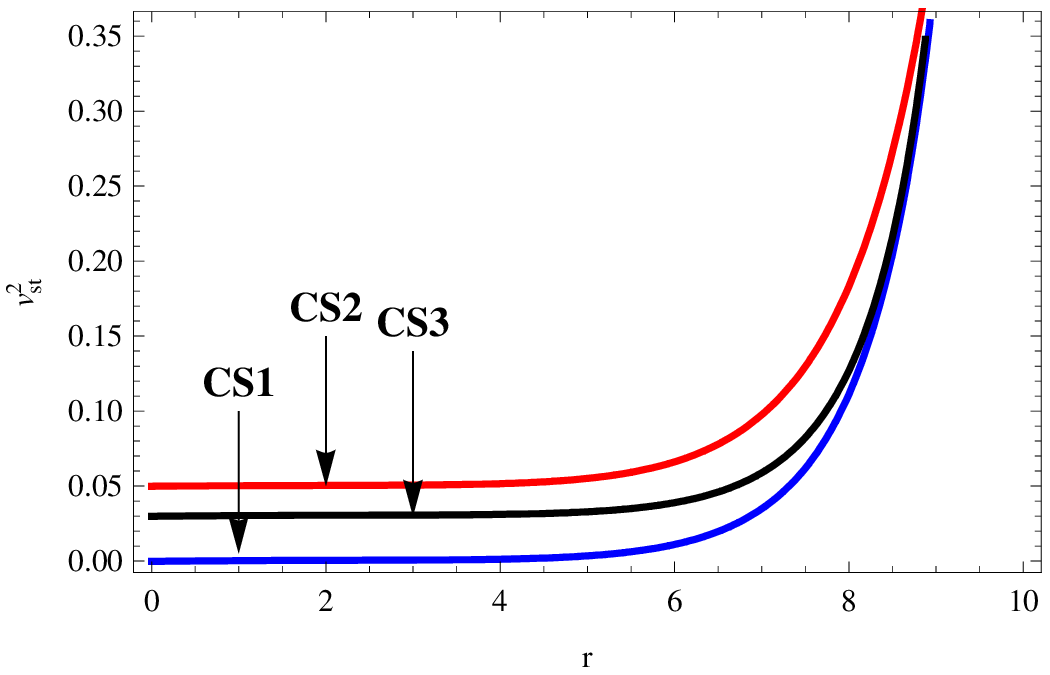}\\
\vspace{2 mm}
\textbf{Fig.14} Variations of $v_{st}^{2}$ versus $r$ (km) with the numerical values of $A$, $B$ and $C$ from Table 2, Table 3 and Table 4 respectively.\\
\vspace{4 mm}

\end{figure}

\begin{figure}

\includegraphics[height=1.35in]{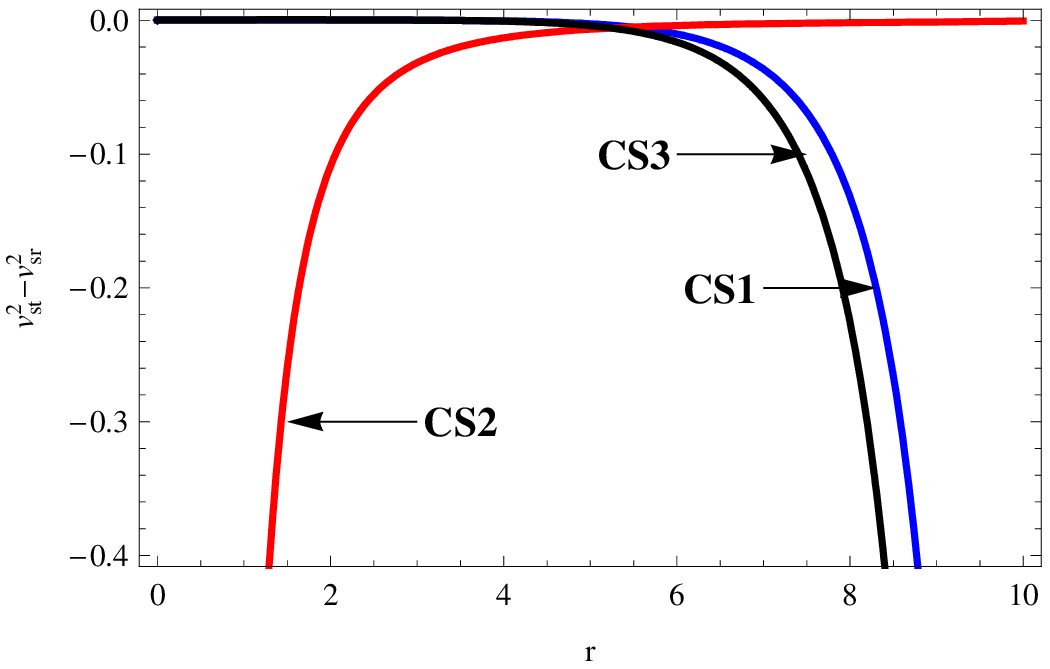}~~
\includegraphics[height=1.35in]{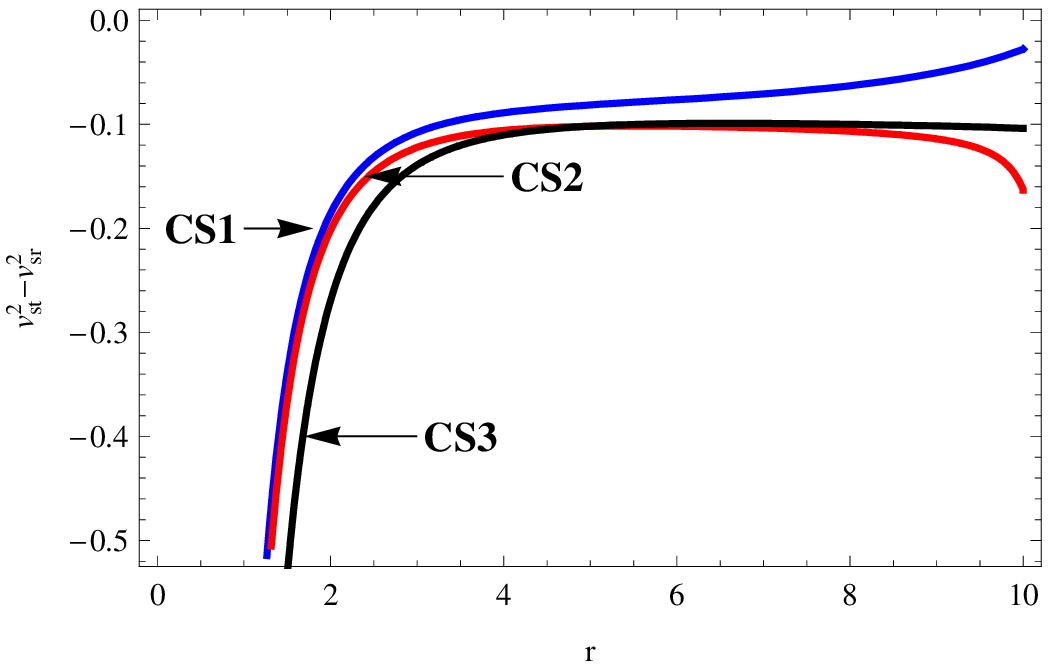}~~
\includegraphics[height=1.35in]{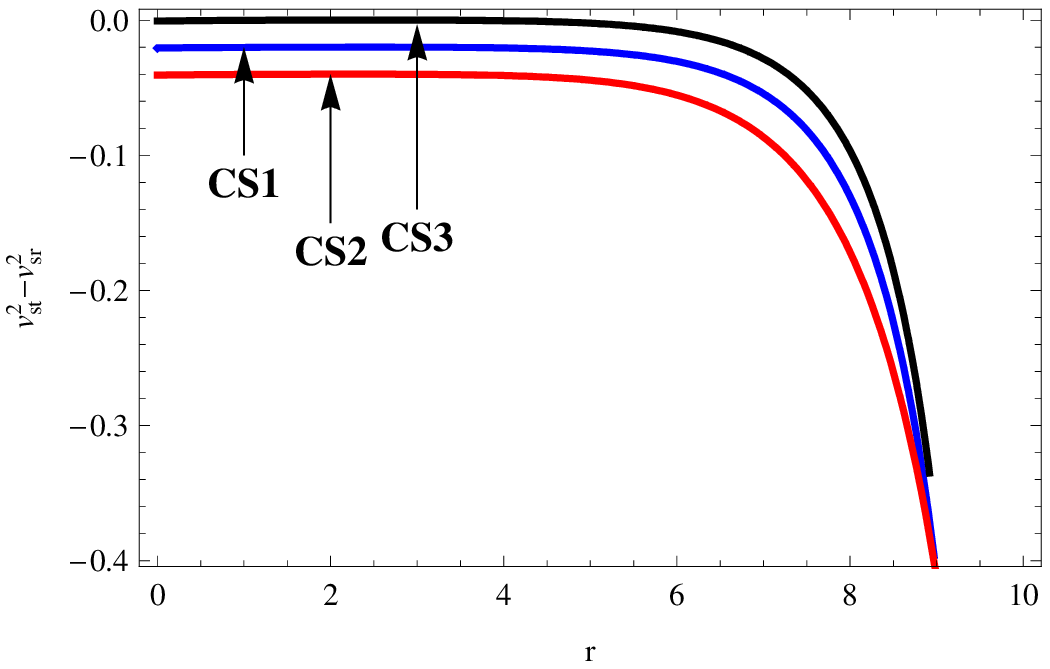}\\
\vspace{2 mm}
\textbf{Fig.15} Variations of $v_{st}^{2}-v_{sr}^{2}$ versus $r$ (km) with the numerical values of $A$, $B$ and $C$ from Table 2, Table 3 and Table 4 respectively.\\
\vspace{4 mm}

\end{figure}

\begin{figure}

~~\includegraphics[height=1.35in]{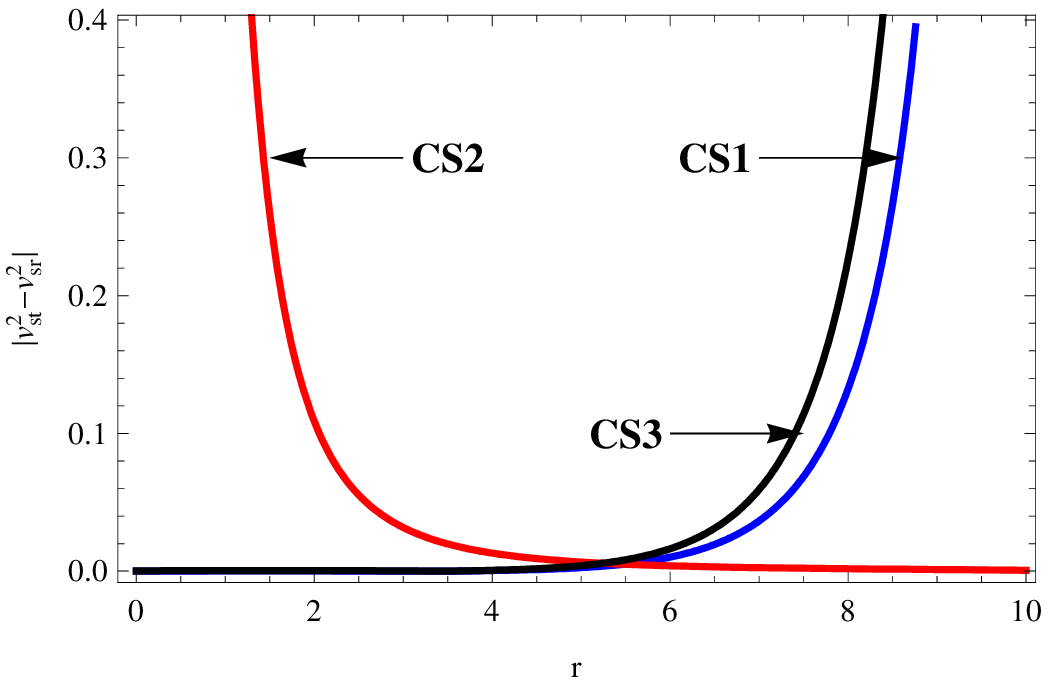}~~
\includegraphics[height=1.35in]{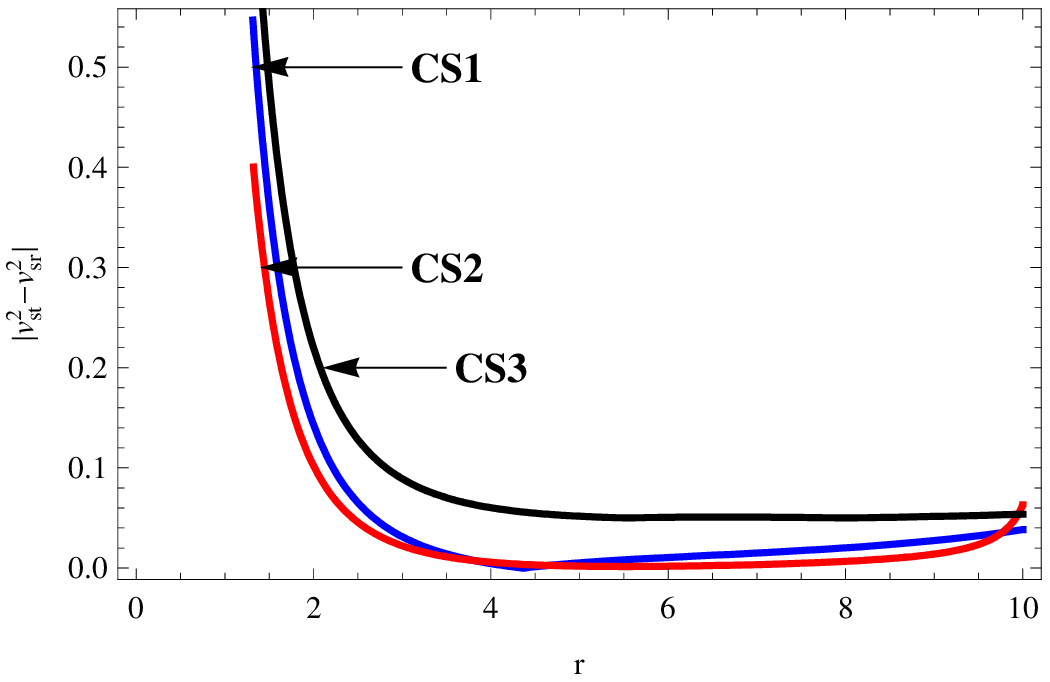}~~
\includegraphics[height=1.35in]{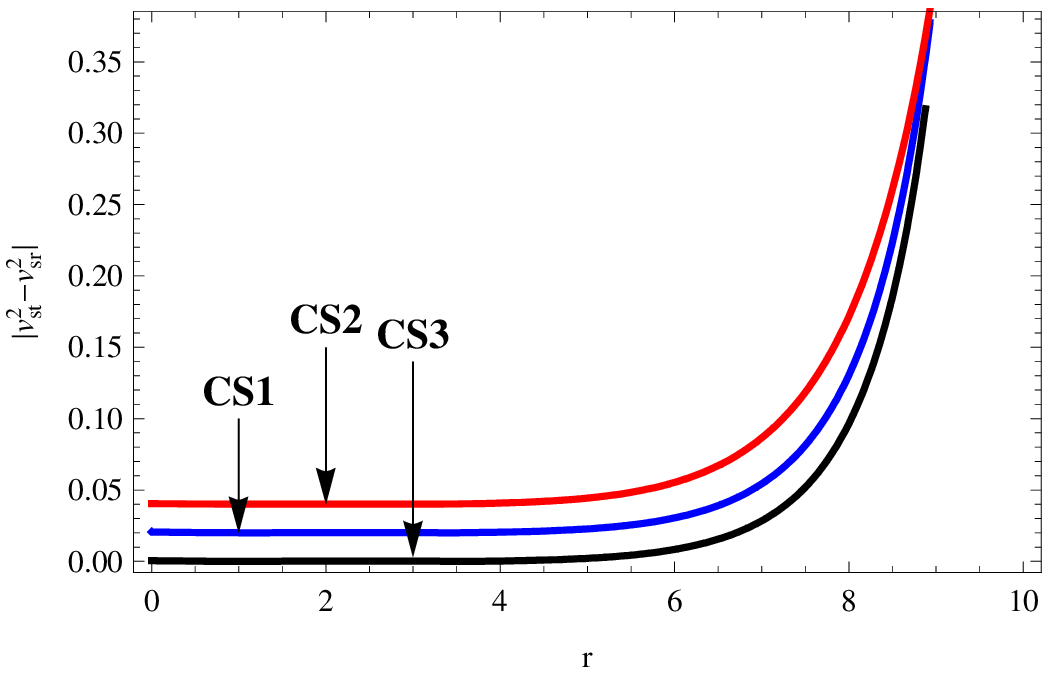}\\
\vspace{2 mm}
\textbf{Fig.16} Variations of $|v_{st}^{2}-v_{sr}^{2}|$ versus $r$ (km) with the numerical values of $A$, $B$ and $C$ from Table 2, Table 3 and Table 4 respectively.\\
\vspace{4 mm}

\end{figure}

\begin{figure}

~~\includegraphics[height=1.35in]{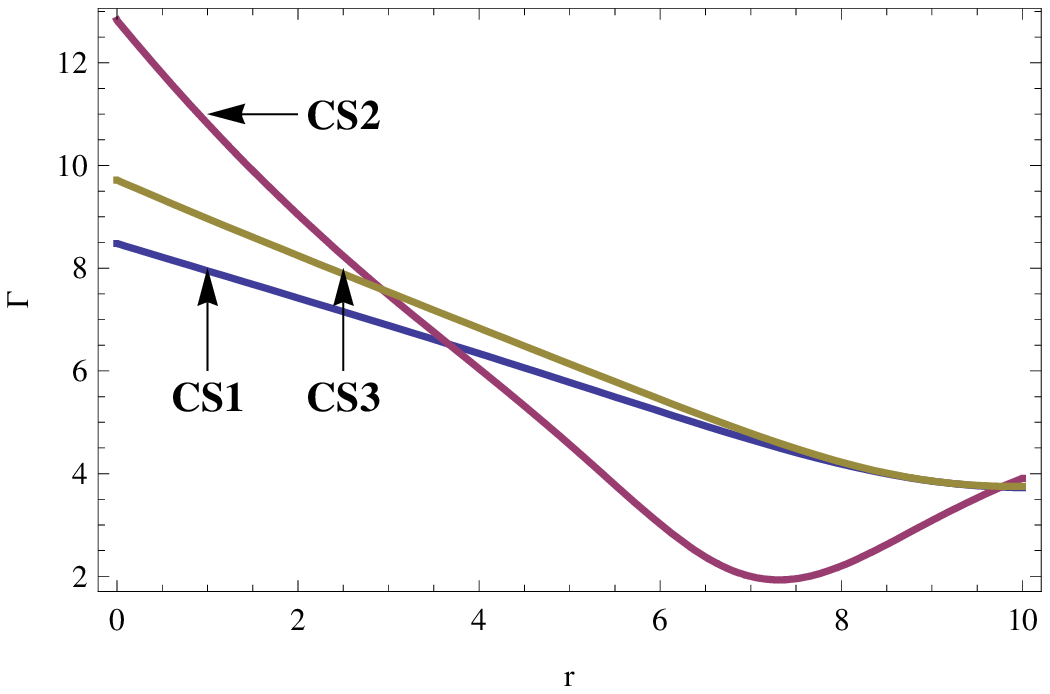}~~
\includegraphics[height=1.35in]{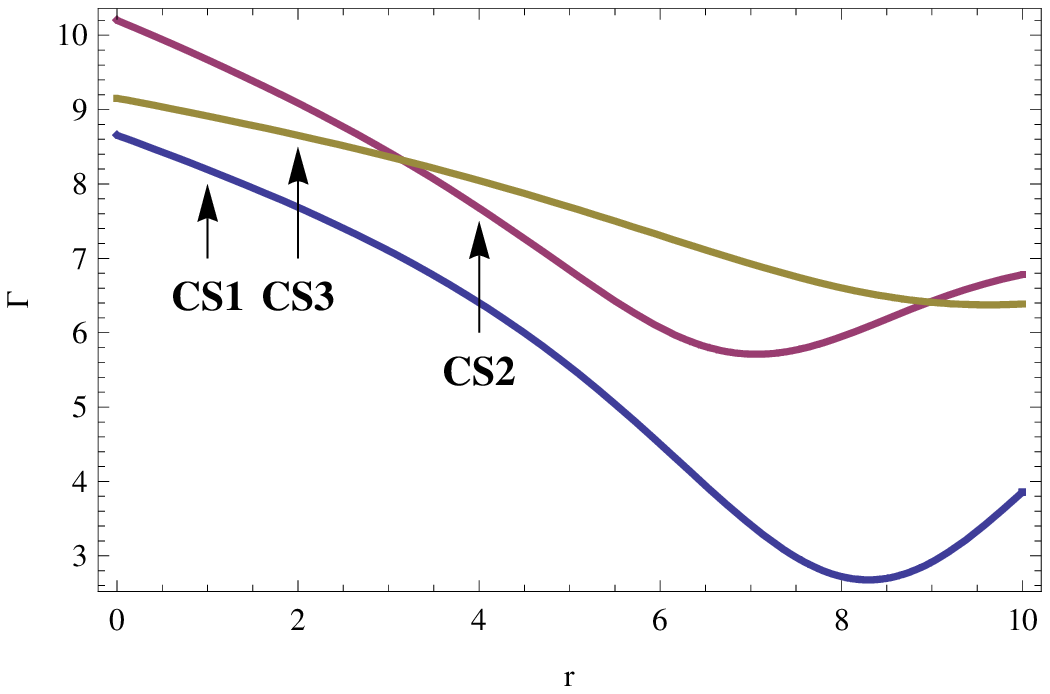}~~
\includegraphics[height=1.35in]{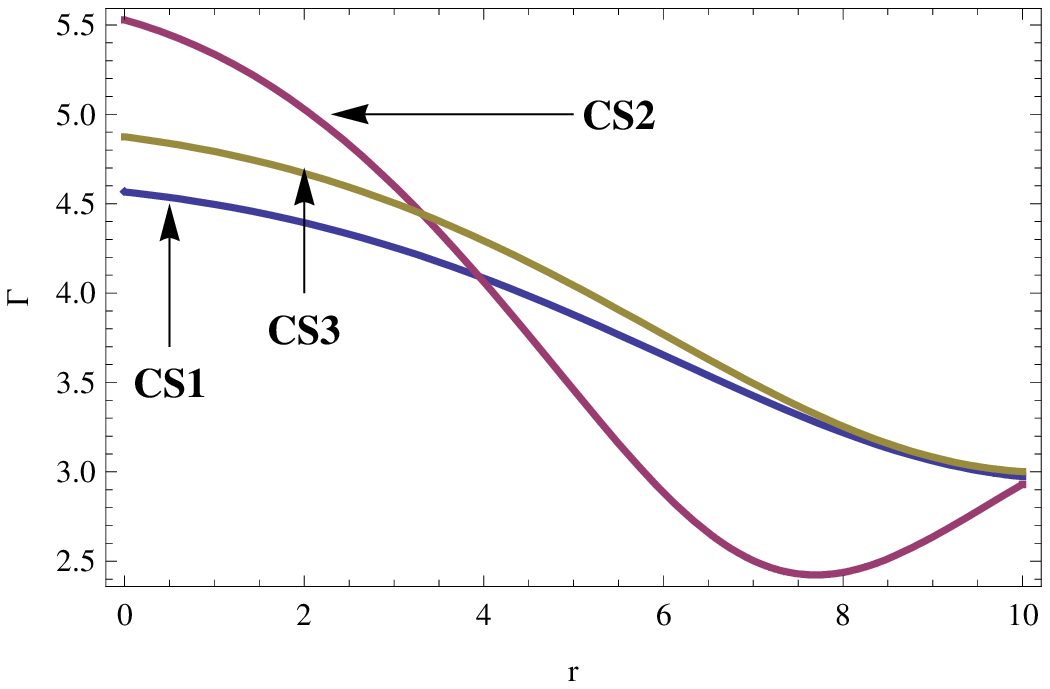}\\
\vspace{2 mm}\\
\textbf{Fig.17} Variations of $\Gamma$ versus $r$ (km) with the numerical values of $A$, $B$ and $C$ from Table 2, Table 3 and Table 4 respectively.\\
\vspace{4 mm}

\end{figure}

\subsection{Effective Mass and Compactness:}
According to \cite{1BP14}, the mass function within the radius $r$ is defined by
\begin{equation}\label{33}
\left.
\begin{array}{ll}
m(r)=\int_{0}^{r} 4\pi r^{2}\rho dr\\
~~~~~~~~=\frac{1}{4(1+\xi)}\int_{0}^{r}\Big[\beta_{1}e^{-Ar^{\gamma}}(\alpha Br^{\alpha}+\beta Cr^{\beta}+\gamma Ar^{\gamma})\\
~~~~~~~~+\sqrt{\beta_{1}^{2}e^{-2Ar^{\gamma}}(\alpha Br^{\alpha}+\beta Cr^{\beta}+\gamma Ar^{\gamma})^{2}+256\pi^{2}r^{4}\zeta(1+\xi)}\Big]dr.
\end{array}
\right.
\end{equation}
Now, we draw the mass functions with respect to radius from the numerical values of Table 2, 3 and 4 respectively. As $r\rightarrow 0$, $m(r)\rightarrow 0$ in Fig.17 for three different compact stars so we can conclude that the mass function is regular at origin. We also notice that the mass function is monotonic increasing with respect to radius.

\begin{figure}

\includegraphics[height=1.35in]{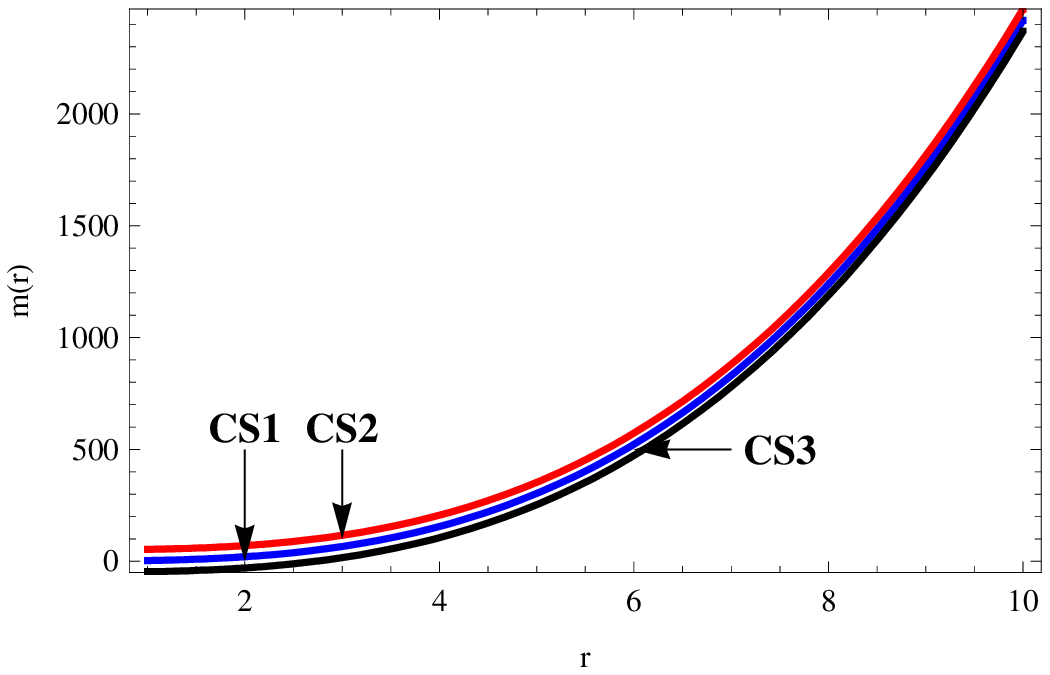}~~
\includegraphics[height=1.35in]{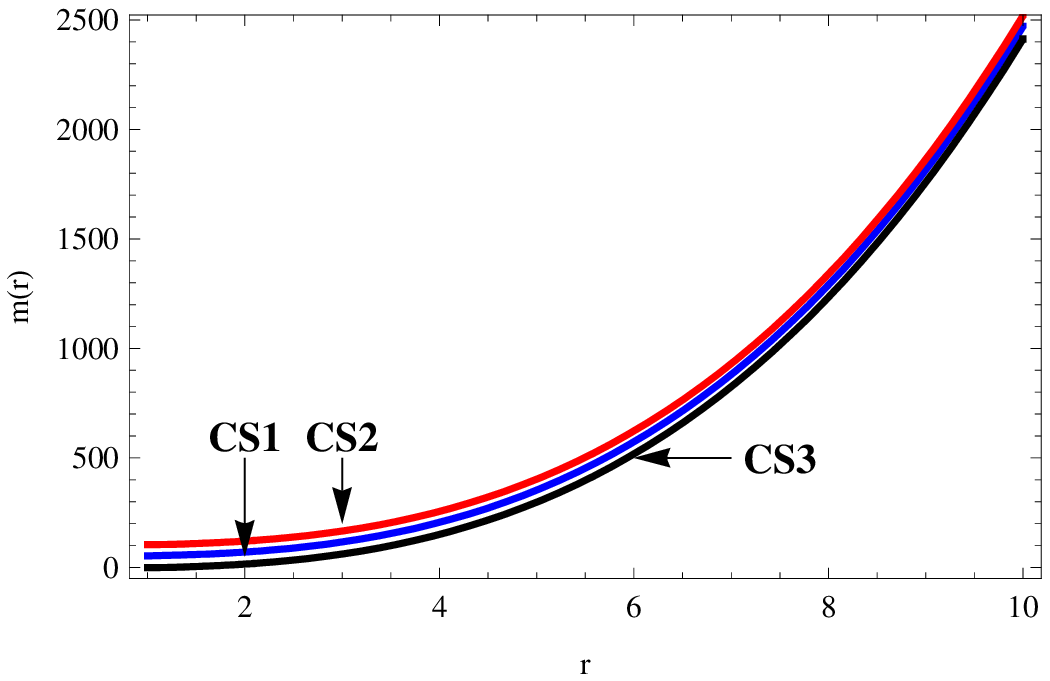}~~
\includegraphics[height=1.35in]{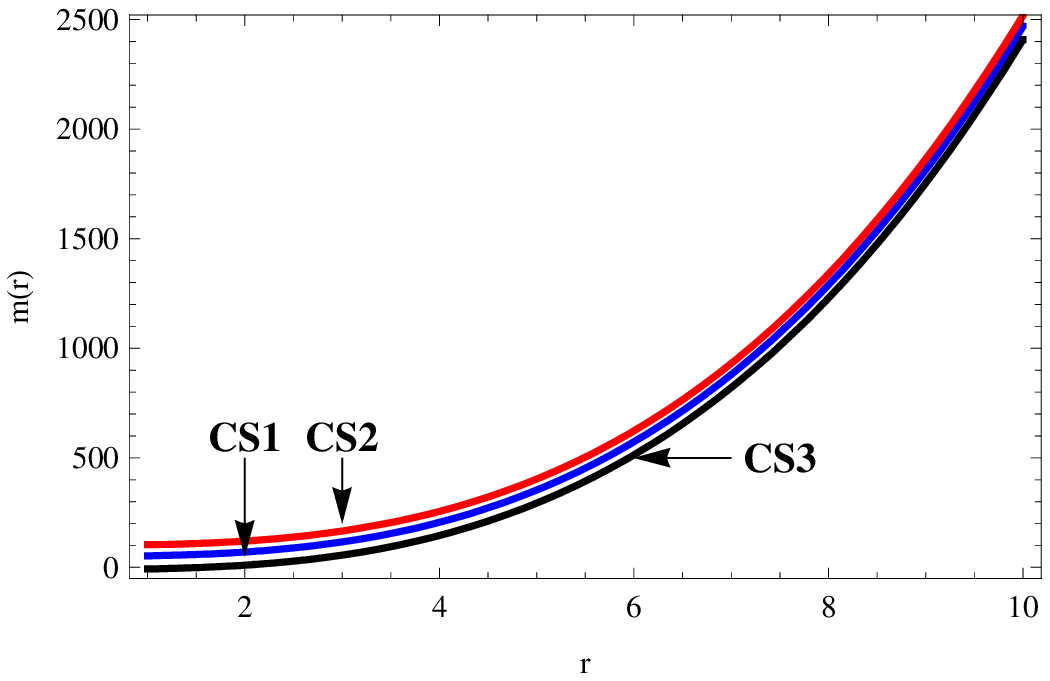}\\
\vspace{2 mm}
\textbf{Fig.18} Variations of $m(r)$ versus $r$ (km) with the numerical values of $A$, $B$ and $C$ from Table 1, Table 2 and Table 3 respectively.\\
\vspace{4 mm}

\end{figure}

Again, the form of compactness of the star is defined by $u(r)$ \cite{1BP14}
\begin{equation}\label{34}
\left.
\begin{array}{ll}
u(r)=\frac{m(r)}{r}\\
~~~~~~~=\frac{1}{4(1+\xi)r}\int_{0}^{r}\Big[\beta_{1}e^{-Ar^{\gamma}}(\alpha Br^{\alpha}+\beta Cr^{\beta}+\gamma Ar^{\gamma})\\
~~~~~~~~+\sqrt{\beta_{1}^{2}e^{-2Ar^{\gamma}}(\alpha Br^{\alpha}+\beta Cr^{\beta}+\gamma Ar^{\gamma})^{2}+256\pi^{2}r^{4}\zeta(1+\xi)}\Big]dr.
\end{array}
\right.
\end{equation}

Now, we evaluate the values of mass, central energy density, central radial pressure and radial pressure at the boundary for the different three compact stars from our model to compare with observational data (See Table 5-7).

\begin{figure}

\includegraphics[height=1.35in]{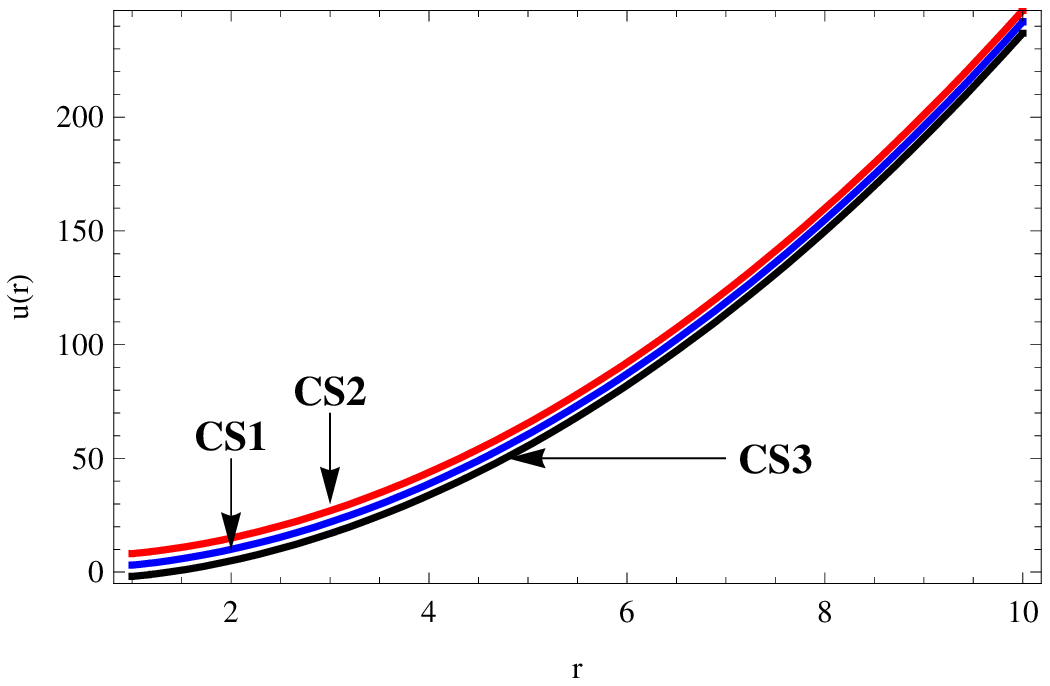}~~
\includegraphics[height=1.35in]{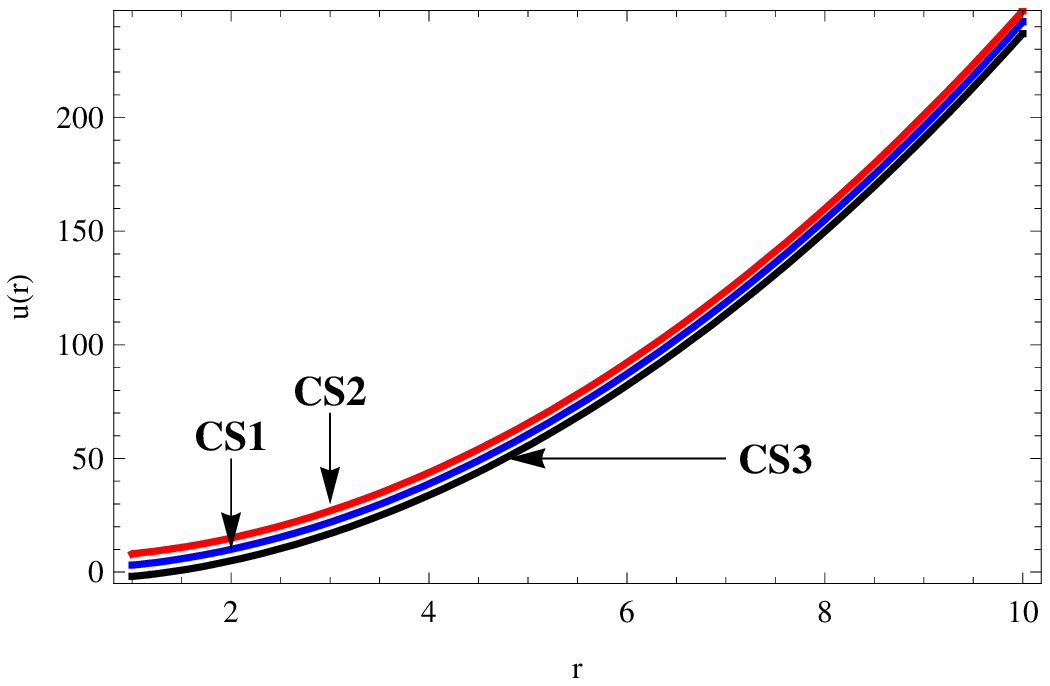}~~
\includegraphics[height=1.35in]{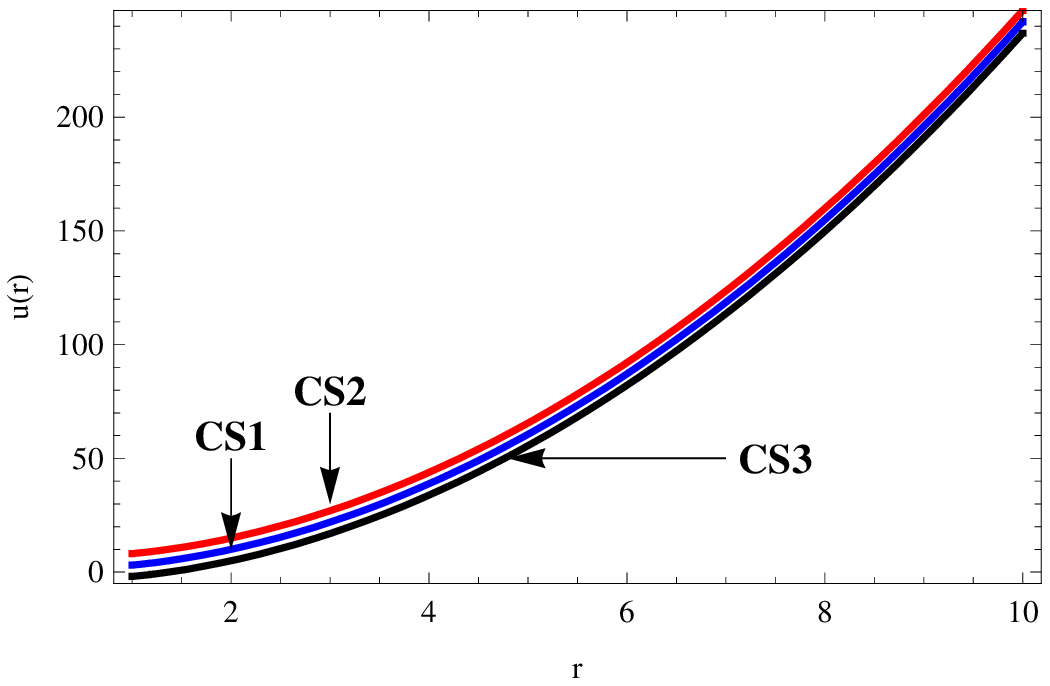}\\
\vspace{2 mm}
\textbf{Fig.19} Variations of $u(r)$ versus $r$ (km) with the numerical values of $A$, $B$ and $C$ from Table 2, Table 3 and Table 4 respectively.\\
\vspace{4 mm}

\end{figure}

\textbf{\begin{table}[h]
\centering
\begin{tabular}{|p{3.9cm}|p{2.5cm}|p{2.2cm}|p{2.7cm}|p{2.6cm}|l|}
\hline
$~~~~~~Compact~Stars$ & $Mass~standard$ & $Mass~from$ & $~~~~\rho_{0} (gm/cc)$ & ~~$\rho_{R} (gm/cc)$ & $~~~p_{0} (dyne/cm^{2})$ \\
~~~~~~~~~~~~~~~~& $data~(in~km)$ & $model~(in~km) $ & ~~~~~~~~~~~~~~~~& ~~~~~~~~~~~~~~~~~~ & ~~~~~~~~~~~~~~~~~~~~\\
\hline
$Vela~X$-$1~(CS1)$ & ~~~~2.61075 & ~~~~2.64376 & 2.403810507$\times10^{15}$ & 2.403792768$\times10^{15}$ & -2.160060323$\times10^{36}$ \\
\hline
$SAXJ1808.4$-$3658~(CS2)$ & ~~~~2.11662 & ~~~~2.10657 & 1.431916338$\times10^{15}$ & 1.431894946$\times10^{15}$ & -1.285972818$\times10^{36}$ \\
\hline
$4U1820$-$30~(CS3)$ & ~~~~3.31875 & ~~~~3.30465 & 1.739754757$\times10^{15}$ & 1.739733365$\times10^{15}$ & -1.562934990$\times10^{36}$ \\
\hline
\end{tabular}
\caption{Calculated values of mass, energy density and pressure of our model from Table 2.}
\end{table}}

\textbf{\begin{table}[h]
\centering
\begin{tabular}{|p{3.9cm}|p{2.5cm}|p{2.2cm}|p{2.7cm}|p{2.6cm}|l|}
\hline
$~~~~~~Compact~Stars$ & $Mass~standard$ & $Mass~from$ & $~~~~\rho_{0} (gm/cc)$ & ~~$\rho_{R} (gm/cc)$ & $~~~p_{0} (dyne/cm^{2})$ \\
~~~~~~~~~~~~~~~~& $data~(in~km)$ & $model~(in~km) $ & ~~~~~~~~~~~~~~~~& ~~~~~~~~~~~~~~~~~~\\
\hline
$Vela~X$-$1~(CS1)$ & ~~~~2.61075 & ~~~~2.60238 & 1.245098292$\times10^{15}$ & 1.245174948$\times10^{15}$ & -1.118544296$\times10^{36}$ \\
\hline
$SAXJ1808.4$-$3658~(CS2)$ & ~~~~2.11662 & ~~~~2.12999 & 1.563552689$\times10^{15}$ & 1.563723828$\times10^{15}$ & -1.404243397$\times10^{36}$ \\
\hline
$4U1820$-$30~(CS3)$ & ~~~~3.31875 & ~~~~3.31554 & 1.328755055$\times10^{15}$ & 1.328878061$\times10^{15}$ & -1.193639739$\times10^{36}$ \\
\hline
\end{tabular}
\caption{Calculated values of mass, energy density and pressure of our model from Table 3.}
\end{table}}

\textbf{\begin{table}[h]
\centering
\begin{tabular}{|p{3.9cm}|p{2.5cm}|p{2.2cm}|p{2.7cm}|p{2.6cm}|l|}
\hline
$~~~~~~Compact~Stars$ & $Mass~standard$ & $Mass~from$ & $~~~~\rho_{0} (gm/cc)$ & ~~$\rho_{R} (gm/cc)$ & $~~~p_{0} (dyne/cm^{2})$ \\
~~~~~~~~~~~~~~~~& $data~(in~km)$ & $model~(in~km) $ & ~~~~~~~~~~~~~~~~& ~~~~~~~~~~~~~~~~~~\\
\hline
$Vela~X$-$1~(CS1)$ & ~~~~2.61075 & ~~~~2.64586 & 1.455574550$\times10^{15}$ & 1.455603073$\times10^{15}$ & -1.307867618$\times10^{36}$ \\
\hline
$SAXJ1808.4$-$3658~(CS2)$ & ~~~~2.11662 & ~~~~2.10746 & 1.627389393$\times10^{15}$ & 1.627433961$\times10^{15}$ & -1.461994045$\times10^{36}$ \\
\hline
$4U1820$-$30~(CS3)$ & ~~~~3.31875 & ~~~~3.31585 & 1.662385577$\times10^{15}$ & 1.662423014$\times10^{15}$ & -1.493684884$\times10^{36}$ \\
\hline
\end{tabular}
\caption{Calculated values of mass, energy density and pressure of our model from Table 4.}
\end{table}}

\begin{figure}
\includegraphics[height=1.35in]{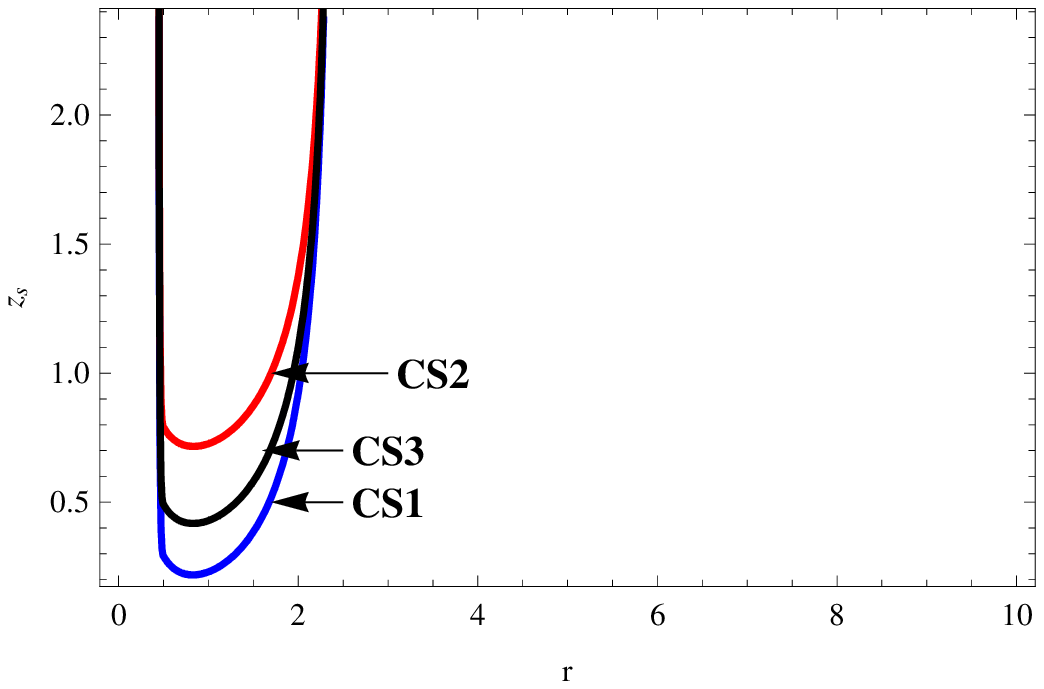}~~
\includegraphics[height=1.35in]{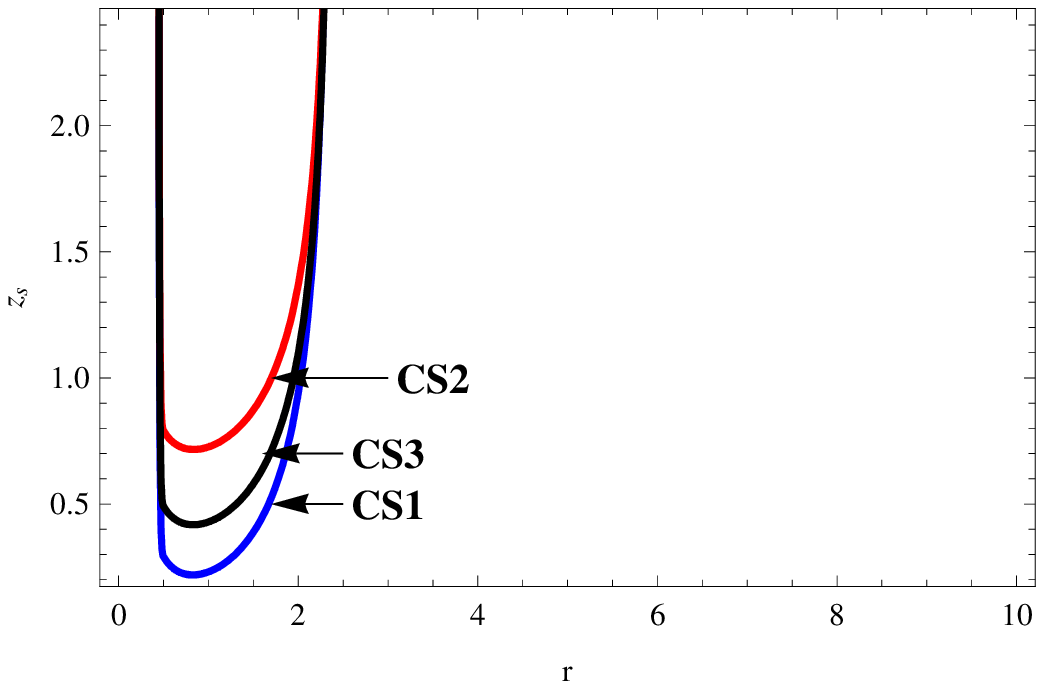}~~
\includegraphics[height=1.35in]{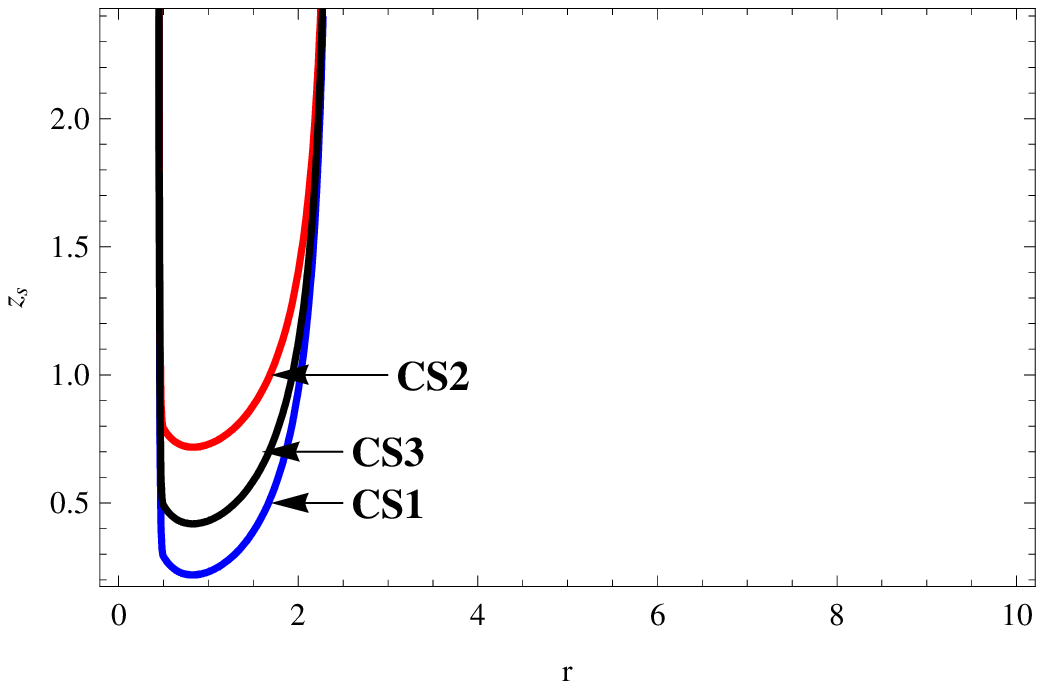}\\
\vspace{2 mm}
\textbf{Fig.20} Variations of $z_{s}$ versus $r$ (km) with the numerical values of $A$, $B$ and $C$ from Table 2, Table 3 and Table 4 respectively.\\
\vspace{4 mm}
\end{figure}

\subsection{Relation between Mass and Radius:}

In this section, we study the relation between mass and radius for three different compact stars to check whether all data are lying in the desired range or not. The factor ``$M/R$" is called compactification factor. We conclude from Table 8-10 that it lies between $1/4$ and $1/2$ \cite{KJTR06}. According to \cite{BHA59}, twice of the compactification factor ($2M/R$) takes maximum allowed value $8/9$ for our model (See Table 8-10).

\subsection{Surface Redshift:}

The redshift function is defined as \cite{1BP14,AGQSJA15,AG15}
\begin{equation}\label{35}
z_{s}=\frac{1}{\sqrt{1-\frac{2m(r)}{r}}}-1.
\end{equation}

We calculate the values of redshift for CS1, CS2 and CS3 in Tables 8-10. According to Bohmer and Harko \cite{1BCGHT06}, the surface redshift can be arbitrarily large, it must be less than $\leq 5$ for an anisotropic star in the appearance of a cosmological constant. Though our model does not contain cosmological constant but the maximum surface redshift from our model is always $\leq 5$ (See Table 8-10 and Fig. 20). So, our anisotropic quintessence compact star model is quite reasonable.\\

\begin{table}
\centering
\begin{tabular}{|l|l|l|l|l|l|}
\hline
$~~~~~~Compact~Stars$ & $\frac{M}{R} (standard~data)$ & $\frac{M}{R}~from~model$ & $~~\frac{2M}{R}<\frac{8}{9}$ & $~~~~\frac{\rho_{0}}{\rho_{R}}$ & $~~~~z_{s}$ \\
\hline
$Vela~X$-$1~(CS1)$ & ~~~~0.273091 & ~~~~0.276544 & 0.553088 & 1.000007416 & 0.495853 \\
\hline
$SAXJ1808.4$-$3658~(CS2)$ & ~~~~0.299381 & ~~~~0.297989 & 0.595918 & 1.000014940 & 0.573132 \\
\hline
$4U1820$-$30~(CS3)$ & ~~~~0.331875 & ~~~~0.330465 & 0.660930 & 1.000012296 & 0.717336 \\
\hline
\end{tabular}
\caption{Calculated values of the desired parameters of our model
from Table 2.}
\end{table}

\begin{table}
\centering
\begin{tabular}{|l|l|l|l|l|l|}
\hline
$~~~~~~Compact~Stars$ & $\frac{M}{R} (standard~data)$ & $\frac{M}{R}~from~model$ & $~~\frac{2M}{R}<\frac{8}{9}$ & $~~~~\frac{\rho_{0}}{\rho_{R}}$ & $~~~~z_{s}$ \\
\hline
$Vela~X$-$1~(CS1)$ & ~~~~0.273091 & ~~~~0.272215 & 0.544431 & 0.999938437 & 0.481572 \\
\hline
$SAXJ1808.4$-$3658~(CS2)$ & ~~~~0.299381 & ~~~~0.301272 & 0.602543 & 0.999890556 & 0.586189 \\
\hline
$4U1820$-$30~(CS3)$ & ~~~~0.331875 & ~~~~0.331554 & 0.663108 & 0.999907436 & 0.722878 \\
\hline
\end{tabular}
\caption{Calculated values of the desired parameters of our model
from Table 3.}
\end{table}

\begin{table}
\centering
\begin{tabular}{|l|l|l|l|p{2.4cm}|l|}
\hline
$~~~~~~Compact~Stars$ & $\frac{M}{R} (standard~data)$ & $\frac{M}{R}~from~model$ & $~~\frac{2M}{R}<\frac{8}{9}$ & $~~~~~~~\frac{\rho_{0}}{\rho_{R}}$ & $~~~~z_{s}$ \\
\hline
$Vela~X$-$1~(CS1)$ & ~~~~0.273091 & ~~~~0.276764 & 0.553527 & 0.999980404 & 0.496589 \\
\hline
$SAXJ1808.4$-$3658~(CS2)$ & ~~~~0.299381 & ~~~~0.298085 & 0.596170 & 0.999972614 & 0.573622 \\
\hline
$4U1820$-$30~(CS3)$ & ~~~~0.331875 & ~~~~0.331585 & 0.663170 & 0.999977480 & 0.723037 \\
\hline
\end{tabular}
\caption{Calculated values of the desired parameters of our model
from Table 4.}
\end{table}

\section{Conclusions}

This work has given out the quintessence dark energy behavior of the anisotropic compact star model in $f(T)$ gravity with modified Chaplygin gas consisting of ordinary matter together with quintessence field along tangential component. Using the diagonal tetrad field we have obtained the equations of motion by taking of quintessence field and electromagnetic field with respect to spherically symmetric metric. We have taken the forms of $a(r)$ and $b(r)$ as $a(r)=Br^{\alpha}+Cr^{\beta}$ and $b(r)=Ar^{\gamma}$ where we have chosen three sets values of these parameters i.e., $\alpha=2, \beta=\gamma=3$; $\alpha=3, \beta=2, \gamma=4$ and $\alpha=4, \beta=2, \gamma=3$. Here, we have assumed the total charge in the form $q(r)=q_{0}r^{m}$. Next, we have solved all of these equations to get $\rho$, $p_{r}$, $\rho_{q}$ and $p_{t}$ in terms of $r$ and some constants. Using matching condition, we have evaluated the values of $A$, $B$ and $C$ in Table 2-4 with respect to Table 1 for those three cases of the parameters $\alpha$, $\beta$ and $\gamma$. By these values of constants we have plotted the figures for the above mentioned physical quantities for CS1, CS2 and CS3 (See Figs.1-5). From these figures we can conclude that our model corresponds to quintessence dark energy model as energy density ($\rho$) is positive, radial pressure ($p_{r}$) is highly negative, $\rho+3p_{r}<0$, energy density corresponding to quintessence field ($\rho_{q}$) is positive and transversal pressure ($p_{t}$) is negative due to considering quintessence field along tangential component of our model. Anisotropic force has been calculated to see whether this is positive or negative. From Fig.6, we can say that $\Delta<0$ always for CS1, CS2 and CS3 i.e., there exists attractive force like quintessence field to ensure our model to be realistic. Next, we have also noticed from Figs.7-9 that the both the energy density ($\rho$) and radial pressure ($p_{r}$) are monotonic increasing in some region and monotonic decreasing in some places with respect to $r$ and they attain various maximum values for CS1, CS2 and CS3 with respect to three sets values of $\alpha, \beta, \gamma$.\\

From Figs.10-12, we have noticed that all energy conditions except SEC are satisfied for all values of parameters for our proposed model. The last condition is satisfied only for $\beta_{2}<0$. By stability analysis given on the basic of Figs.13-17, we have seen that $0<v_{sr}^{2},v_{st}^{2}\leq 1$, $v_{sr}^{2}>v_{st}^{2}$, $|v_{st}^{2}-v_{sr}^{2}|\leq 1$ and $\Gamma>4/3$ for all of three cases. So, we can guarantee that our model is potentially stable.\\

In Tables 5-7, we have evaluated numerical values of masses of CS1, CS2 and CS3 for our model using equation (37) to manifest the closeness with the observational data. Also, we have calculated the corresponding values of central and surface density and central pressure in these tables. Due to Fig.18, masses of these stars are tending to zero when $r\rightarrow0$. From Tables 8-10, we have calculated the compactification factor ($u_{r}$) using equation (34) and observed that the twice of compactification factor is always less than $<\frac{8}{9}$. We have plotted the figure of $u(r)$ in Fig.19 which tells that $u(r)\rightarrow0$ when $r\rightarrow0$. In these tables, using equation (35), we have also calculated the numerical values of surface redshift and have drawn figure in Fig.20. From both calculation and figure, we have maximum values of the surface redshift function which is always less than $5$. So, our proposed anisotropic compact stars model in $f(T)$ gravity with quintessence field and modified Chaplygin gas is completely rational.\\

\end{document}